\documentclass[english,11pt]{article}
\usepackage{babel}
\usepackage{filecontents}
\usepackage[T1]{fontenc}
\usepackage[latin9]{inputenc}
\usepackage{geometry}
\usepackage{color}
\usepackage{placeins,array,multirow,tablefootnote,floatpag,booktabs,makecell,longtable}
\floatpagestyle{plain}
\usepackage{float} 
\usepackage{nicefrac}
\usepackage{caption}
\usepackage[flushleft]{threeparttable}
\usepackage{amstext}
\usepackage{amsthm}
\usepackage{graphicx}
\usepackage{enumitem}
\usepackage{titlesec}
\usepackage[labelformat=simple]{subcaption}

\titleformat{\subsubsection}[runin]{\bfseries}{}{}{}[]
\usepackage[authoryear]{natbib}
\usepackage{overpic}
\usepackage{xcolor}
\newcommand\hl[1]{%
  \bgroup
  \hskip0pt\color{black} #1%
  \egroup
}
\usepackage{xr-hyper}
\makeatletter
\newcommand*{\addFileDependency}[1]{
  \typeout{(#1)}
  \@addtofilelist{#1}
  \IfFileExists{#1}{}{\typeout{No file #1.}}
}
\makeatother

\newcommand*{\myexternaldocument}[2][]{%
    \externaldocument[#1]{#2}%
    \addFileDependency{#2.tex}%
    \addFileDependency{#2.aux}%
}

\myexternaldocument[app:]{spSIR-appendix}
\theoremstyle{plain}

\providecommand{\empiricalname}{Empirical Implication}
\newcommand{\notegrowth}{\emph{Note:} Growth rate of Infected (left panel) and Infected at date $t$ as a fraction of the population (right panel). }
\newcommand{\notelegend}{The legend of the right panel indicates the steady state fraction of Recovered.}
\newcommand{\notelegendbis}{The legend inside each panel indicates the steady-state fraction of Recovered.}
\newcommand{\notemix}[4]{\emph{Note:} Infected at date $t$ as a fraction of the population#1 #2 #3 #4}

\usepackage[unicode=true,pdfusetitle,
 bookmarks=false,bookmarksnumbered=true,bookmarksopen=false,bookmarksopenlevel=2,
 breaklinks=true,pdfborder={0 0 0},pdfborderstyle={},backref=page,colorlinks=true]
 {hyperref}
\hypersetup{allcolors=darkblue}
\definecolor{darkblue}{rgb}{0,0.1,0.65}

\renewcommand*{\backref}[1]{}
\renewcommand*{\backrefalt}[4]{{\ifcase #1%
          \or(Cited on page~#2)%
          \else(Cited on pages #2)%
    \fi%
    }}

\begin{document}

\setcounter{footnote}{3}
\author{
    Alberto Bisin 
    \and 
    Andrea Moro}

\title{Learning Epidemiology by Doing: \\ The Empirical Implications of a Spatial-SIR Model with Behavioral Responses\thanks{Bisin: New York University, \href{https://wp.nyu.edu/albertobisin/}{\texttt{wp.nyu.edu/albertobisin/}}, 
\texttt{alberto.bisin@nyu.edu}. Moro: Vanderbilt University, \href{https://andreamoro.net}{\texttt{andreamoro.net}}, \texttt{andrea@andreamoro.net}. We thank Pedro Sant'Anna,  Giorgio Topa,  Maxim  Pinkovskiy, the editor and two anonymous referees for their helpful
comments on earlier drafts of this paper, and Gianluca Violante for suggestions about the calibration. }
}
\date{\today \\ {\small First version: June 10, 2020}}
\maketitle
\begin{abstract}
    We simulate a spatial behavioral model of the diffusion of an infection to understand the role of geographic characteristics: the number and distribution of outbreaks, population size, density, and agents' movements.
    We show that several invariance properties of the SIR model concerning these variables do not hold when agents interact with neighbors in a (two dimensional) geographical space. Indeed, the spatial model's local interactions generate matching frictions and local herd immunity effects, which play a fundamental role in the infection dynamics. We also show that geographical factors change how behavioral responses affect the epidemic. We derive relevant implications for estimating the effects of the epidemic and policy interventions that use panel data from several geographical units. 
\end{abstract}

\newpage

\section{Introduction}

The SARS-CoV-2 epidemic has diffused at very different rates across countries and cities.\footnote{See  \cite{Desmet2020} and  \cite{fernandez2020estimating}.} Plots of case statistics over time by location 
are common with media and opinion leaders to compare the dynamics of the epidemic across geographical units, often intending to evaluate the effect of different policy interventions.  But, how can we compare the United States to Ireland or New York to Miami, given their differences in population size, density, and other geographic and socio-economic characteristics?  How do we export parameter estimates about the epidemic obtained from the city of Vo', a small town near Padua, in Italy, or from the Diamond Princess cruise ship,  to understand the diffusion of the epidemic in New York city?\footnote{See 
\cite{Lavezzo-Vo-Study} and \cite{Mizumoto_2020} 
respectively.} 

This paper proposes a spatial model of epidemic diffusion, the Spatial-SIR model, to study how the dynamics of an epidemic scales in relevant geographical characteristics: the number and distribution of outbreaks, population size, density, and agents' movements.\footnote{
    \cite{duranton-puga} overview the importance of geographic factors in economic modeling.
    } 
To this end, we adopt an Agent-Based methodology, that is,  we simulate populations of agents characterized by micro-level rules of behavior over time and space.\footnote{See, e.g., \cite{el2012social, hunter2018comparison} for surveys and methodological discussions. More generally, for an introduction and survey of Agent-Based models in the social sciences, see, e.g.,  \cite{billari2012agent, bruch2015agent}. For spatially explicit  Agent-Based models in the social epidemiology tradition of detailed forecasting models, see
\cite{dunham2005agent, grefenstette2013fred}; and \cite{hunter2017taxonomy} for a survey.}  \label{epiresearch} Research in epidemiology has extended SIR, allowing for detailed descriptions of the geographic, socio-economic, and demographic characteristics of the population of interest and its environment. In social epidemiology, Agent-Based models generally incorporate detailed, granular  assumptions on human behavior and the social and physical environment, with the aim of forecasting with accuracy and precision the dynamics of an epidemic (as, say, meteorological models of weather dynamics).\footnote{\hl{See, e.g., \cite{eubank2004modelling}, the research at  \href{https://covid19.gleamproject.org}{GLEAM project}, 
\href{https://www.mobs-lab.org/projects.html}{mobs-lab}, and the  \href{https://www.imperial.ac.uk/mrc-global-infectious-disease-analysis}{Imperial college MRC Centre for Global Infectious Disease Analysis} available, respectively, at https://covid19.gleamproject.org, https://www.mobs-lab.org/projects.html, and https://www.imperial.ac.uk/mrc-global-infectious-disease-analysis.}}  In this paper instead we aim at identifying the stylized effects of geographic characteristics on the dynamics emerging from an abstract spatial-SIR  model.

The Spatial-SIR model imposes restrictions on the epidemic dynamics that depend on each location's geographical characteristics, informing comparisons across locations. We show that it is not possible to uncover these restrictions from 
the workhorse model or epidemic diffusion, the SIR model (\cite{kermack1927contribution, kermack1932contributions}). These restrictions are consequential
for empirical analysis using longitudinal infection data. 
Most of the recent wealth of contributions to the study of the SARS-CoV-2 epidemic in economics has  restricted its epidemiology component to SIR and does  not account
for the geographic characteristics that we focus on in this paper.\footnote{\hl{See e.g., \cite{atkeson2020will}, \cite{eichenbaum2020macroeconomics},  \cite{brotherhood2020economic}, and \cite{jarosch2020internal}}}
Several  exceptions introduce spatial dimensions to SIR, but focus on how connections between  geographical units affect the epidemic diffusion.\footnote{
\hl{See, e.g., 
\cite{argente2020cost},
\cite{antras2020globalization}, \cite{birge2020controlling},  \cite{bognanni2020economic},
\cite{brady2020fragmented},
\cite{cua2020structural}, \cite{fajgelbaum2020optimal}, \cite{giannone}, \cite{glaeser2020much}.}} In this paper instead, we focus on the comparative dynamics of the epidemic  with respect to different geographical characteristics of (closed) units. The spatial dimensions we account for in the present paper introduce local interactions in the contact process between agents.\footnote{\hl{Relatedly, \cite{acemoglu2020testing}, \cite{alfaro2020social},  \cite{Azzimontietal2020}  extend SIR to explicitly model the epidemic dynamics in networks; \cite{ellison} allows for heterogeneity of the contact process between subpopulations.}}

Section \ref{invariance} begins by highlighting the relevant invariance properties of SIR with respect to the geographic characteristics we focus on in this paper. Section \ref{sec:spatialSIRtheory}  introduces the Spatial-SIR model. Individuals
are placed in a two-dimensional
space and travel in this space at a given speed. When infected, they
can only infect their neighbors with a probability that we
interpret as the strength of the virus. 
Spatial-SIR determines the diffusion rate of infection 
depending on epidemiological and geographic factors that are 
confounded in one parameter of the standard SIR.
Section \ref{sec:SIRcomp} shows how distinguishing these factors is crucial in Spatial-SIR because the local interactions arising in the model generate matching frictions across agents. What we
call ``local herd 
immunities'' arise from the constrained movement of people
in space. Local herd immunities break several  invariance relationships which hold in the SIR model (enumerated in Section  \ref{invariance}), where
susceptible individuals match with infected individuals randomly.

Section \ref{sec:spatialSIR} presents several simulations of a calibrated  Spatial-SIR to study the roles of geographic characteristics. Section \ref{sec:behavioral} incorporates behavioral responses, not accounted for in the SIR model, into Spatial-SIR to highlight how their effects on the infection diffusion depend on geographic factors.%
\footnote{ \hl{
Models of rational agents limiting contacts to reduce the risk of  infection are relatively scarce in epidemiology; \cite{fenichel2013economic} \cite{weitz2020moving} are prominent examples, see also \cite{funk2010modelling} and  \cite{verelst2016behavioural}  for surveys. Most importantly, the formal modeling of behavioral responses has not yet broken into the large forecasting models which represent the core of the discipline as e.g., \cite{balcan2009multiscale}, \cite{balcan2010modeling}, \cite{chinazzi2020effect}, and   \cite{ferguson2020report}. 
Not surprisingly, behavioral responses are instead central to epidemiological models in economics. Early contributions in this respect include \cite{geoffard1996rational} and  \cite{goenka2012infectious}; while recent work includes 
\cite{acemoglu2020multi}, 
\cite{aguirre}, \cite{argente2020cost}, 
\cite{bethune2020covid}, \cite{farboodi2020}, \cite{fernandez2020estimating}, 
\cite{gans2020},
\cite{greenwood2019equilibrium},
\cite{keppo2020behavioral},   \cite{toxvaerd2020equilibrium}, as well as several of the papers cited above modeling spatial extensions of SIR. See also \cite{BisinMoroRat2020} for a formal introduction to SIR with forward-looking rational-choice agents. None of these papers discuss the effects of the interaction between behavioral and spatial factors in the spread of an infection which we show has important implications for the dynamics of herd immunity and possibly for the effects of Non-Pharmaceutical Interventions (NPI). } 
} 

Section \ref{sec:empirical} presents five implications for 
empirical analysis we learn from our model simulations. We note that research exploiting geographic variation in longitudinal data 
to study the effect of policy interventions or other covariates on the epidemic
outcomes must deal with 
time-varying heterogeneity across locations that is hard to control 
for without imposing a specific structure. 
\hl{
We simulate data from Spatial-SIR models to highlight how structural estimates of SIR models fail to incorporate restrictions imposed by the spatial structure. We quantitatively assess the implied error of the effect of a lockdown policy that this mis-specification generates.

Reduced-form methods  exploiting the different time and location of the  implementation of the intervention have been adopted to study the effect of policy interventions or other covariates on the epidemic
outcomes, often using a Difference-in-differences design.\footnote{\hl{Without any attempt at being exhaustive, see, e.g.,   
\cite{allcott2020polarization},
\cite{allcott2020explains},
\cite{bartik2020measuring},
\cite{borri2020great},
\cite{borsati2020questioning},
\cite{breen2021distributional}
\cite{Chernozhukov2020},
\cite{chetty2020economic},
\cite{couture},
\cite{courtemanche2020did}, 
\cite{courtemanche2020strong},
\cite{crucini2020stay},
\cite{dave2020jue},
\cite{fang2020human},
\cite{gapen2020assessing},
\cite{glaeser2020learning},
\cite{goolsbee2020fear},
\cite{gupta2020effects},
\cite{hsiang2020effect},
\cite{juranek2020effect},
\cite{kong2020disentangling},
\cite{maloney2020determinants},  \cite{mangrum2020college}, \cite{pepe2020covid},
\cite{sears2020we},
\cite{wilson2021weather},
\cite{wright2020poverty},
\cite{yilmazkuday2020covid},
\cite{ziedan2020effects}.
} 
}
We run Difference-in-differences regressions on the simulated data to highlight how such an approach fails to capture the heterogeneity of the effects by time and locations. Perhaps not surprisingly, it also fails to identify the average treatment effect of the policy if the data does not include both treated and untreated units at locations with different geographic characteristics, or if treatment occurs at different times in different locations.\footnote{\cite{bacon2020} describes some threats to the validity of
DiD-design in the analysis of NPIs to fight
the spread of COVID-19. See also \cite{CALLAWAY2020} and \cite{callaway2021policy} for approaches to mitigate the problems arising from these threats.}}

Our paper highlights that empirical research can gain from 
imposing the cross-location restrictions implied by the epidemiological models. Spatial and time heterogeneity render the reduced-form analysis of these data problematic. Short of directly estimating structural models, simulated data can be exploited in an integrated structural/reduced-form empirical analysis.

\section{Invariances in the SIR Model  \label{invariance} }

We first introduce the standard SIR model as a benchmark to evaluate the role of adding spatial structure. The society is populated by $N$ agents that are ex-ante identical. Let $\mathcal{S}=\{S, I, R\}$ denote the individual state-space, indicating Susceptibles, Infected, and Recovered.\footnote{In \cite{bisinmoro2020} we expand the state space to better capture some relevant aspects of the SARS-CoV-2 epidemic by adding Symptomatics and Dead. This expansion of the state space is inconsequential for studying the effects of geographical characteristics but adds realism, helping to study its policy implications.} 
Let $h_t=[S_t,I_t,R_t]$ denote the distribution of the population across the state-space at time $t$. 
The following transitions govern the dynamics of $h_t$: i) a Susceptible agent becomes infected
 upon contact with an infected, with probability $\beta {I_t}/{N}$; ii) an  agent infected at $t$, can recover at any future period with probability $\rho$; 
iii) a Recovered agent never leaves this state
(this assumes that Recovered agents are immune to infection). 

The SIR can be solved analytically.\footnote{See e.g., \cite{herben42hethcote}, \cite{Moll2020},  \cite{neumeyer2020clase}.} The equations describing its dynamics in discrete time are  
\begin{equation}
   \Delta  I_t = \beta S_t \frac{I_t}{N} - \rho I_t, \; \; \; \Delta R_t=\rho I_t, \; \; \; S_t+I_t+R_t=N.  \label{idot}
\end{equation} 
Parameter $\beta$ is to be interpreted as the infection rate and is related to $\mathcal{R}_0= \beta/\rho$, the number of agents a single infected agent infects, on average, at an initial condition $R_0=0, \; I_0 > 0$. The infection rate $\beta$ can be decomposed  as the infection rate per-contact between a Susceptible and an Infected, $\pi$, and the number of contacts per unit of time, $c$:   $\beta=\pi c$ (in the continuous time limit). Distinguishing
the roles of the number of contacts  and  the contagion rate is conceptually 
important to avoid interpreting  $\mathcal{R}_0$ and  $\beta$ as structural parameters of the model. In the Spatial-SIR we introduce in the next section, they are the result of virological,
geographical and, in Section \ref{sec:behavioral}, behavioral factors.

We highlight three invariance properties of the dynamics of the SIR model, with respect to initial conditions 
and to the spatial and virological parameters driving the dynamics. 
We study the robustness of these invariances to the introduction of a spatial structure. 
\newline

\noindent {\bf Stationary state invariance to initial conditions.} 
Given any $I_0>0$,\footnote{Initial conditions are uniquely represented by $I_0$, since $R_0=0$ and $S_0=N-I_0$.} the dynamic system converges to a  unique stationary state. Namely, the size of the initial outbreak, $I_0$,  does not affect the stationary state.  
This stationary state of Infected is $I_{\infty}=0$, while the stationary state of Recovered (the fraction of the population infected in the course of the epidemic), $0<{R_{\infty}}/{N}<1$,  is characterized uniquely in terms of $\mathcal{R}_{0}=\beta/\rho$, as the solution of the following fixed point
equation:
\begin{equation}
\frac{R_{\infty}}{N}=-\frac{1}{\mathcal{R}_{0}}\ln(1-\frac{R_{\infty}}{N}).\label{ss}
\end{equation}

\noindent {\bf Transitional dynamics invariance to initial conditions (in the limit ${I_0}/{N} \rightarrow 0$).} The   dynamics of ${h_t}/{N}$ depends on initial conditions only via  ${I_0}/{N}$. It is then  invariant as the fraction of infected at the initial condition converges to zero, ${I_0}/{N}  \rightarrow 0$. In particular, the peak of infected cases in this limit is 
\begin{equation} 
    \frac{I^{peak}}{N}=1-\frac{1}{\mathcal{R}_0} \left(1+\log \mathcal{R}_0 \right). 
    \label{peak} 
\end{equation}

\noindent {\bf Transitional dynamics invariance to contacts  and probability of contagion,  keeping $\beta$ constant.} 
The  dynamics of ${h_t}/{N}$ depends on the number of contacts  $c$  and probability of contagion, $\pi$, but is invariant to changes in  $c$  and $\pi$  that leave $\beta=\pi c$ constant. \newline 

\label{transdyn}If the epidemic is governed by SIR, these invariances provide restrictions of the model which are testable with cross-city data i) when different cities have different initial conditions (infection outbreaks) $I_0$; and/or ii) when differences in the number of contacts and in the probability of contagion map into differences in  $\beta=\pi c$ and $\rho$ across cities.\footnote{Variation across virological characteristics can, in principle, be studied with data across different epidemic. In this paper, we concentrate mainly on geographical variation across cities.} 

\section{The Spatial-SIR model} \label{sec:spatialSIRtheory}

We add a spatial dimension to SIR by locating agents in a 2-dimensional space, which we call the ``city.'' Agents are ex-ante identical in terms
of demographics and symmetric in terms of location.
Agents are randomly located initially, and every day $t=[0,T]$, they travel distance $\mu$ 
toward a random direction. \label{travelspeed}By doing so, they potentially meet new individuals every day, therefore $\mu$ is an abstraction of the speed in which agents find new contacts, potentially in a different state than their previous neighbors.
Two agents come into contact when they are at a geographical distance closer than $p$.\footnote{\label{spatialb}The spatial behavior of agents as postulated in Spatial-SIR  is mechanical and it abstract from the network structure, e.g., home, work, city, which characterizes real world behavior. This is to focus more clearly and directly on highlighting the fundamental effects of spatial behavior in determining the dynamics of the epidemic, as well as their stylized dependence on various  geographical characteristics (outbreaks, population size, density, agents' movements); see \cite{bisinmoro2020} for an extension of Spatial-SIR allowing for a network structure.} 

Spatial-SIR is represented by the following transitions: 
i) a Susceptible agent in a location within distance $p$ from the location of an Infected becomes infected with
probability $\pi$;
ii) an Infected agent can Recover at any period with probability $\rho$;  
iii) Recovered agents never leave these states. 
The resulting dynamical system is difficult to characterize formally.\footnote{\label{reac-diff}In Appendix \ref{app:sec:app_theory} we write it as a Markov chain on configurations in space, along the lines of interacting particle-system models \citep{ kindermann1980american, liggett2012interacting}. Some properties are obtained by analogy to the physics of percolation on lattices; see  \cite{grassberger1983critical}, \cite{tome2010critical}. For local-interaction models in Economics see  \cite{blume2011identification}, \cite{conley2007estimating}, \cite{glaeser2001measuring},  \cite{ozgur2011dynamic}. In Appendix \ref{app:sec:app_theory} we also discuss the mathematical formulation of this class of models in continuous time and space, as reaction-diffusion equations systems (see e.g. \cite{chinviriyasit2010numerical} and \cite{wu2017epidemic}).}
We turn then to simulations. 

Table \ref{parameters} reports the calibrated parameters we use in the baseline model. We calibrate transitions between states, $[S, I, R]$, to various SARS-CoV-2 parameters from epidemiological studies, notably, e.g., \cite{ferguson2020report}. We calibrate $\beta$ (in its components $\pi$ and $c$) and the agents' daily travel
distance $\mu$ to data on average contacts from \cite{Mossong_2008} and to match estimates of initial (prior to policy interventions) growth rates of the epidemic in Lombardy, Italy.\footnote{We acknowledge the substantial uncertainty in the literature with respect to even the main epidemiological parameters pertaining to this epidemic. As we noted in the introduction, this is less damaging when aiming at understanding mechanisms and orders-of-magnitude rather than at precise forecasts.} The calibration is performed as follows. 
 \begin{table}
 	\thisfloatpagestyle{empty}
	\caption{Calibrated parameter values: baseline model}\label{parameters}  
    \centering
    \begin{tabular}{clcc}
        \toprule
      &  Parameter                       & Notation  & Value\tabularnewline
        \midrule
     (1) &   number or people                & $N$       & 26,600\tabularnewline
     (2) &  initially infected              & $I_0$    & 30 \tabularnewline
     (3) &   prob. of recovery               & $\rho$    & 0.154   \tabularnewline 
      (4) &  average contacts per day                & $c$       & 13.5 \tabularnewline
     (5) &  contagion radius                & $p$       & 0.013\tabularnewline
     (6) &  contagion probability           & $\pi$     & 0.054 \tabularnewline
     (7) &  mean distance traveled          & $\mu$     & 0.034 \tabularnewline
        \bottomrule
    \end{tabular}
\end{table}

\subsubsection*{(1)-(2) Population geography, initial conditions}	
We choose $N$ so that our simulations converge in a reasonable time (see the beginning of 
	Section \ref{sec:outbreaks} for a description of how the model scales in size). We place people      initially on a unit square
	drawing their $x$ and $y$ coordinates independently from a Uniform distribution 
	$\sim U[0,1]$%
	\footnote{In the simulation with heterogeneous density we set initial locations at a distance 
	from the center drawn randomly from a Normal distribution $\sim N(0,1)$ and direction
	drawn from a Uniform distribution  $\sim U[0,2\pi]$.}.
	At all $t>0$, individuals 
	are relocated at distance $\mu$ from their location at $t-1$, in a direction randomly
	drawn  from a Uniform distribution  $\sim U[0,2\pi]$. When individuals get close to the boundary, 
	the movement is constrained to point to a direction opposite to the boundary. At time $t=0$ we  
	set 30 individuals in Infected state; all 
	others are Susceptibles. In all specification excepts those reported in Figures  \ref{fig:City-cluster-comparisons} (\ref{fig:cluster-rates}) and  \ref{fig:hetdensity} the 
	Infected at $t= 0$ are those initially located closest to location $
	[x=0.25,y=0.25]$. 

\subsubsection*{(3) Transition away from the infected state, $I$.}  

    The probability 
	 any agents transitions away from state $I$ is $\rho$, hence the average time an agents stays in state  $I$ is 
	 $T_\text{inf}={1}/{\rho}$. 
	 We set $\rho$ to match a theoretical moment which holds exactly at the initial condition in SIR. Recall $\mathcal{R}_{0}$ denotes the number of agents a single infected agent at $t=0$ 
	 infects, on average. Let $g_0$ 
 	denote the growth rate of the number of infected agents at $t=0$. Then, in SIR, 
	${(\mathcal{R}_{0}-1)}/{T_\text{inf}}=g_0$\
	for $t \rightarrow 0$. 
	For SARS-CoV-2, $\mathcal{R}_{0}$ is reasonably estimated
	between $2.5$ and $3.5$. (\cite{huang2020clinical},
	\cite{remuzzi2020covid},
	\cite{zhang2020estimation},
	\cite{paules2020coronavirus}). 
	The daily rate of growth of infections $g$ is estimated to be between $0.35$ and $0.15$
	by \cite{kaplan2020gianluca}, \cite{alvarez2020simple}, and \cite{ferguson2020report}. This implies, from the equation above for $g_0$, that $T_\text{inf}$ is between 4 and 7 days
	(respectively for $\mathcal{R}_{0}$ between 2.5 and 3.5).  \cite{ferguson2020report} use 6.5 days, which we use to set $\rho=1/6.5$. 
	
	\subsubsection*{(4)-(5) Contagion circle radius.} The contagion radius, $p$, is not separately identified from the average number of contacts, $c$. We set it to match the 
	the average number of contacts observed in demographic surveys. \citet{Mossong_2008} suggests an average of  13.5 contacts every day. 
	 \subsubsection*{(6)-(7) Infection and contact rates.} 
	After setting parameters (1)-(5), we calibrate the remaining parameters $\pi$ (hence $\beta= \pi c$) and $\mu$ to match the daily growth rates of the dynamics of infections observed in the first 35 days of epidemic using data for Lombardy, Italy. Since the number of infections is not observed, we match the growth rates of deaths in the data. This is justified when, as we assumed, the case fatality rate is constant, and Death follows infection after a constant lag on average. Appendix Figure \ref{app:fig-appgrowths} illustrates goodness of fit.   


\begin{figure}
    
    \caption{Geographic progression of infections and recoveries, baseline model}

\makebox[\textwidth][c]{        
    \setlength\tabcolsep{0pt}
    \begin{tabular}{ccc}
        \includegraphics[width=0.33\linewidth]{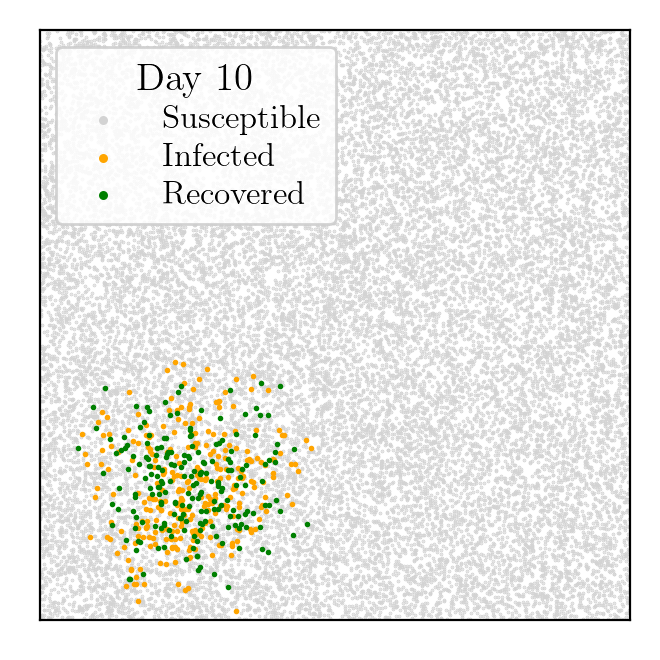} &  
        \includegraphics[width=0.33\linewidth]{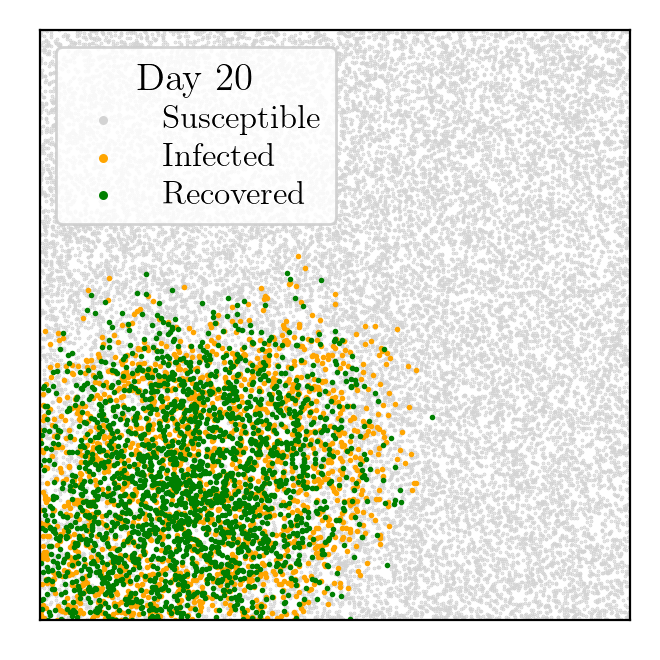} &
        \includegraphics[width=0.33\linewidth]{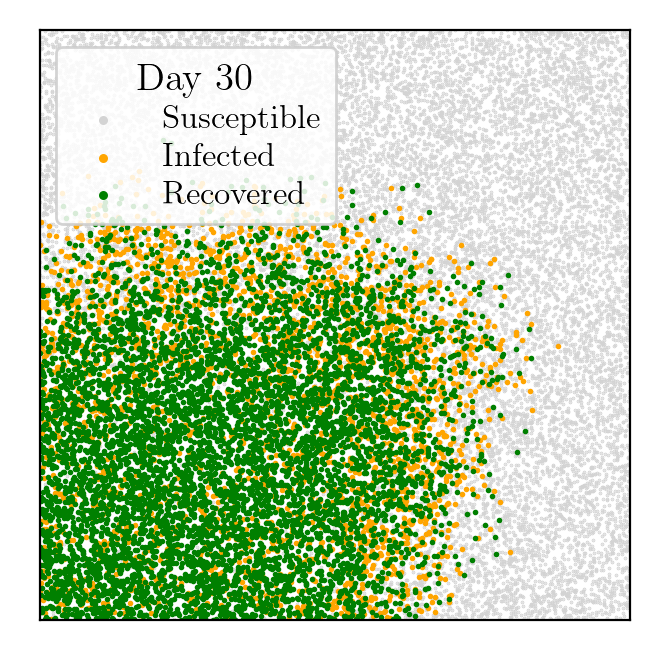}
    \end{tabular}
}
   \label{fig:geo}
   \caption*{\normalfont\footnotesize Note: Position of individuals in the city at day 10, 20, and 30 since the start of the infection, color-coded by the agent's state, model calibrated with baseline parameters}
\end{figure}


The dynamics of an epidemic in the Spatial-SIR model cannot be analyzed in closed form, but Figure \ref{fig:geo} illustrates this dynamics over time and space at the calibrated parameters. The epidemic spreads exponentially from the location of the outbreak.\footnote{All our simulations converge to a unique distribution over the state space $ \left[S,I,R \right] $.
} In the next sections, we compare the dynamics under SIR and Spatial-SIR.   

\section{Local Herd Immunity}\label{sec:SIRcomp}
 
To understand how Spatial-SIR differs from SIR, we simulate the evolution over time of the infection growth rates 
and the fraction
of active cases (that is, infected agents, $I_t/N$) in both models.
We show that geography and people's movements in Spatial-SIR generate local interactions
and matching frictions creating a form of \emph{local herd immunity}, absent in SIR. 
\label{randmatch}Formally, in SIR, random matching implies that  the probability that
any  Susceptible agent is infected at time $t$ is $\beta {I_t}/{N}$.\footnote{
We exploit the continuous-time approximation for ease of exposition. In discrete-time (hence in the simulations), the probability that a susceptible agent is infected per unit of time after $c$ contacts is $1-(1-\pi {I}/{N})^c$.
} 
In Spatial-SIR, this probability is not common across susceptible agents as it depends on the distribution of agents and their states across space, say $H_t$.\footnote{Formally, $H_t$ is a stochastic process whose realization at $t$ maps any agent (or any location, in an equivalent formulation) into a state $h \in \{S, I, R\}.$} We can then describe the probability that a susceptible agent is infected {\em on average} in the Spatial-SIR model as $\beta \lambda(H_t)$, for a well-defined function $\lambda$ which encodes the effects of local herd immunity.\footnote{In Spatial-SIR, the probability that a susceptible agent is infected is not linear in ${I_t}/{N}$ as in SIR. As a consequence, the Spatial-SIR cannot be mapped into a SIR model for some $\beta$ depending on $t$ (see the Appendix for a more formal description of Spatial-SIR). We discuss this point more explicitly when we draw the empirical implications of our analysis in Section \ref{sec:empirical}.}
 
\begin{figure}
        \caption{SIR and Spatial-SIR comparison of infection dynamics}
        \label{fig:densitycomp}
   	\centering
        \includegraphics[width=0.85\linewidth]{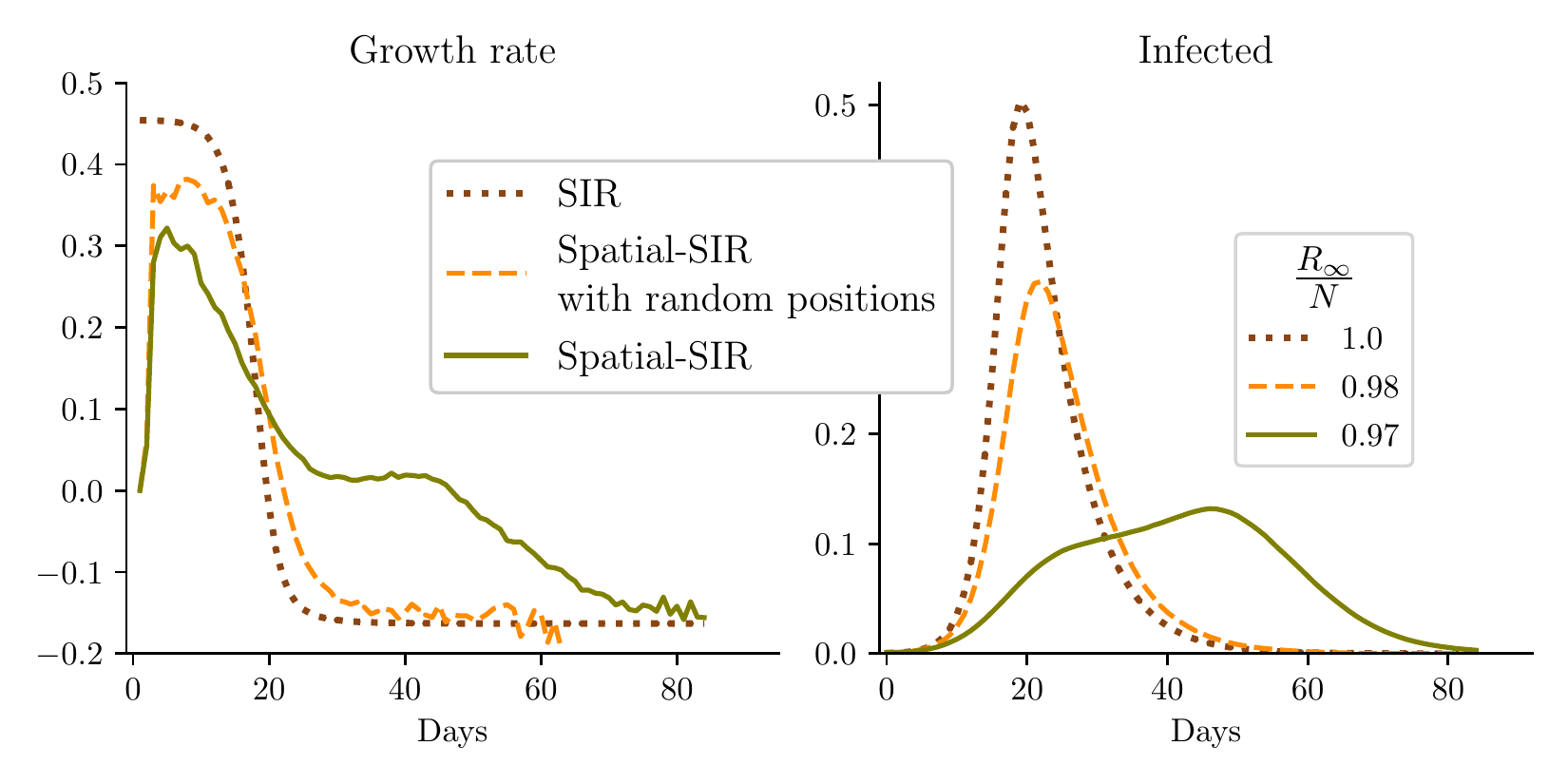}
 
    \caption*{\normalfont\footnotesize 
    \notegrowth
    \notelegend\ Solid green lines: spatial-SIR baseline calibration; dashed orange lines: baseline Spatial-SIR, except with individuals' geographic locations drawn randomly every day with the same rules used to draw initial locations; brown dotted lines:  SIR  with infection and recovery rates equivalent to the ones implied by the baseline calibration of Spatial-SIR.}
\end{figure}

More specifically, in Figure \ref{fig:densitycomp} we compare simulations of Spatial-SIR, 
with the parameters we calibrated for our baseline model, and of SIR, with $\beta$ equal to our calibrated value of the contagion rate multiplied by the average number of daily contacts, implied our calibrated city's population density and contagion radius. 
 The Spatial-SIR displays lower growth rates than SIR initially. ``Local herd immunities'' slow down the diffusion of infection in the early stages and accelerate it afterward (as aggregate herd immunity is delayed); in other words,   $\lambda(H_t)$ is initially smaller and then larger than $I_t/N$. In the figure, the dashed yellow lines report growth rates for a version of Spatial-SIR with agents placed in a random location in the city \emph{every day}, mimicking the random matching aspect of SIR and therefore minimizing the formation of local herd immunities. As expected, the  effect of local herd immunity is much weaker in this model.\footnote{
Local herd immunity is determined by accessibility to susceptible population. Therefore, it is not limited to geography as in our model, but also by the network structure of contacts; see \cite{bisinmoro2020} for a Spatial-SIR model where heterogeneous agents are structured in a network. The role of local herd immunity in both SIR and Spatial-SIR outcomes is also evident when comparing the effects of Non-Pharmaceutical Interventions such as lockdown policies (see Section \ref{sec:empirical}). 
}   

The effect of local herd immunity indicates that the ``reduced-form'' nature of SIR models is missing a potentially important role of matching frictions and, more generally, of local dynamics. Similar considerations can be obtained looking at $\mathcal{R}_0$  (a random variable in Spatial-SIR because the number of contacts of an individual is random).  Replicating simulations of our baseline Spatial-SIR, we estimate \textbf{$\mathcal{R}_0$} as the average number of people infected 
by those infected during the first five days. 
This estimate is within
the range used to calibrate transition rates in many
studies (between 2.5 and 3.5) but is highly volatile. In 20 random replications of the model, the average $\mathcal{R}_0$ is 3.17, with a standard deviation of 0.58.
However, in Spatial-SIR this volatility does not translate into similarly different aggregate outcomes as predicted by standard SIR. The fraction of people ever infected in steady-state averages to 0.97 in the 20 replications, with a standard deviation of 0.001. This suggests that, in our model, $\mathcal{R}_0$ loses its role as the fundamental driving parameter of the epidemic: outcomes are also highly sensitive to characteristics of the
initial cluster of infection.\footnote{Outcomes are also somewhat sensitive to the precise location of the initially infected; therefore the simulation dynamics we display in simulations of Spatial-SIR report averages of 20 random replications of the models}.
While the infection rate during the initial stages is uniquely determined by the structural parameters  $\mathcal{R}_0$ and $\rho$, which are (relatively) independent of the spatial structure of the model,  the infection dynamics rests on the spatial local interaction structure. The growth rate of the infection declines early on following a form of local herd immunity. Indeed, this is what we observe in the data, and we set parameters to match.


\section{Oubreaks,  Population Size, Density, and Agents' Movements} \label{sec:spatialSIR}

In this section, we study the comparative dynamics of the spread of an epidemic as it depends on various relevant geographical characteristics across cities, which determine matching frictions. We compare the effects of these geographical characteristics on both stationary and transitional dynamics in Spatial-SIR with those in SIR. More specifically, we capture the properties of the stationary states by the fraction of recovered, ${R_{\infty}}/{N}$ (as ${I_{\infty}}/{N}=0$ and ${S_{\infty}}/{N}=1-{R_{\infty}}/{N}$). We capture the properties of the transitional dynamics, on the other hand, by the time it takes to for an outbreak to reach the peak of active cases, a measure of the speed of the epidemic, and the height of the peak of active cases as a fraction of the population, ${I^{peak}}/{N}$, a measure of the intensity of the epidemic.

The geographical characteristics we concentrate on are outbreaks,  population size, density, and agents' movements. The number of initial outbreaks and the population size appear directly 
in both Spatial-SIR and SIR. In Spatial-SIR, however, it is the entire distribution of initial outbreaks, not just the number $I_{0}$, which affects the dynamics of the epidemic. As a consequence, the effects of population size in Spatial-SIR are mediated by the initial conditions regarding the distribution of outbreaks. 
City density instead appears directly in Spatial-SIR, but has only an indirect counterpart in SIR.  City density in fact can be thought of as affecting parameter $\beta$ proportionally in SIR, because density affects the number of contacts $c$ by $c=d \Psi$, where $\Psi$  is the contagion area of any (susceptible) individual (which is maintained constant in all simulations), and $\beta=\pi c$.
Finally, agents' movements appears directly in Spatial-SIR, as the average distance traveled every day, 
but is absent in SIR. 

Generally, we show that all these geographic characteristics affect the epidemic differently in Spatial-SIR and SIR. In particular, we highlight that the simulated dynamics of Spatial-SIR do not satisfy some of the invariance properties of the SIR dynamics we have delineated in Section \ref{invariance}. Fundamentally, the effect of local herd immunity on the dynamics is a function of geographic characteristics which, abusing notation, we denote as  $g=(I_0,N, d, \mu)$. We represent this by writing $ \lambda \left( H_t; g\right)$.\footnote{As we noted, the dependence of  $\lambda$ on  $H_t$ changes over time and cannot be expressed in closed form; the same holds for its dependence on $g$. In other words, the dependence of Spatial-SIR on geographic characteristics is structurally distinct from what we would obtain in a SIR model allowing for  $\beta$ to depend on  $g$. We discuss this point more explicitly in Section \ref{sec:empirical}.}  This analysis of the effects of geography on the dynamic of the epidemic has some clearcut implications that empirical cross-city studies of epidemic dynamics should account for; we discuss these in some detail in Section \ref{sec:empirical}.

\subsection{Outbreaks}\label{sec:outbreaks}

The epidemic dynamics in SIR only depends on initial conditions through ${I_0}/{N}$ (first and second invariance in Section \ref{invariance}). This implies scaling in size, that is, the invariance of the dynamics over $x$-times larger cities with $x$-times as many initial infected. 
 In Spatial-SIR, however, as we noted, the space of the initial conditions is larger, as it includes the spatial distribution of infections. Indeed, the spatial distribution matters greatly in Spatial-SIR. Scaling in size obtains in Spatial-SIR only if initial infection outbreaks are appropriately homogeneously distributed across space.\footnote{In Appendix Figure 
\ref{app:fig:2sizes} we compare the  
progression of the contagion between the baseline city 
and a city with four
times the population and the area (so that density is constant), and with four initial clusters of the same size as in the baseline
located in symmetric locations. The progression of the infection is nearly identical, barring minor effects due to the randomness of people's locations and movement.} 

\begin{figure}[!ht]
    
    \caption{Geographic progression of infections and recoveries}\label{fig:clustercomp}
    
\begin{subfigure}{1.\linewidth}
\makebox[\textwidth][c]{ 
    \begin{tabular}{ccc}
        \includegraphics[width=0.33\linewidth]{figuresdir/nc5-baseline_pos-day10.png} &  
        \includegraphics[width=0.33\linewidth]{figuresdir/nc5-baseline_pos-day20.png} &
        \includegraphics[width=0.33\linewidth]{figuresdir/nc5-baseline_pos-day30.png}
    \end{tabular}
}
    \caption{Baseline model} 
    \label{fig:geo2}
\end{subfigure} 

\begin{subfigure}{1.\linewidth}
    \makebox[\textwidth][c]{ 
    \begin{tabular}{ccc}
    \includegraphics[width=0.33\linewidth]{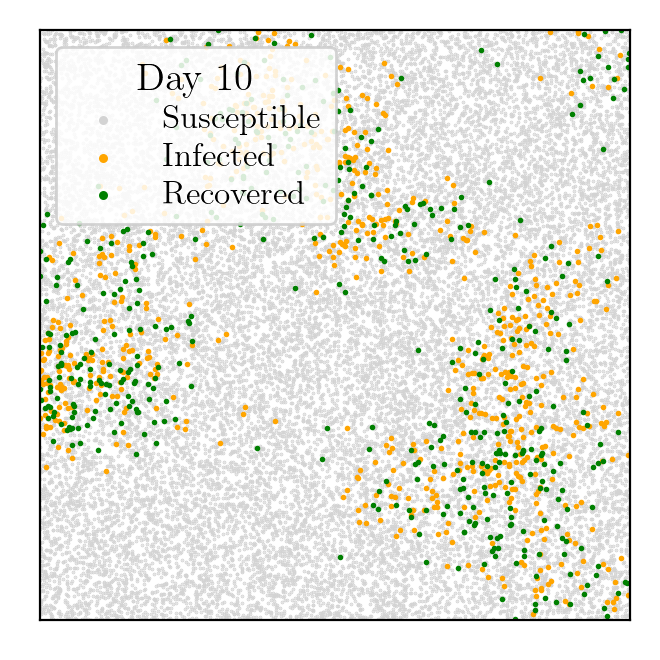} & 
    \includegraphics[width=0.33\linewidth]{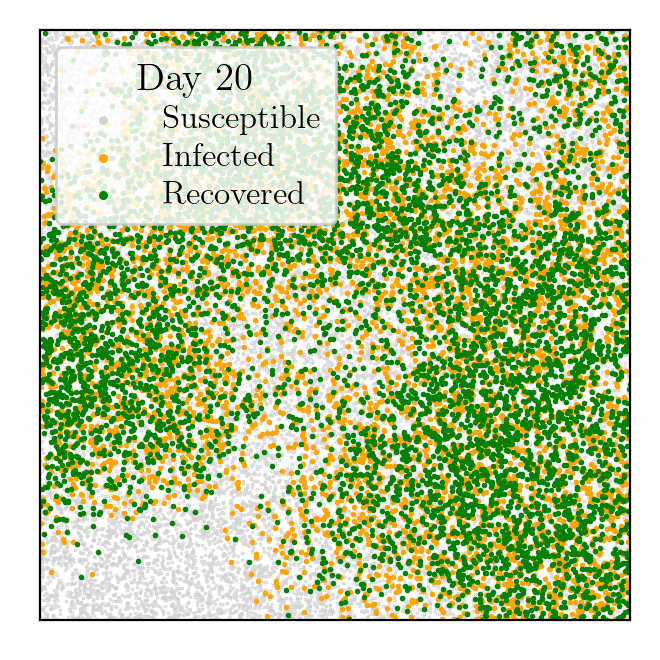} & 
    \includegraphics[width=0.33\linewidth]{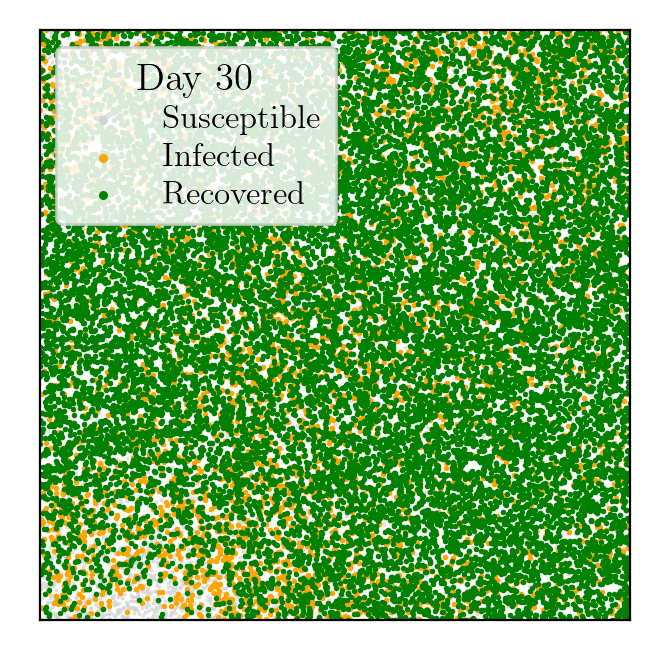}
    \end{tabular}
    }
    \caption{Initial clusters of contagion randomly located}
    \label{fig:City-cluster-comparisons}
\end{subfigure} 

   \caption*{\normalfont\footnotesize Note: Position of individuals in the city at day 10, 20, and 30 since the start of the infection, color-coded by the agent's state}
\end{figure}

\begin{figure}[t]
    \caption{Infection dynamics: inital outbreaks in random locations}\label{fig:cluster-rates}
    \centering
    \includegraphics[width=.85\textwidth]{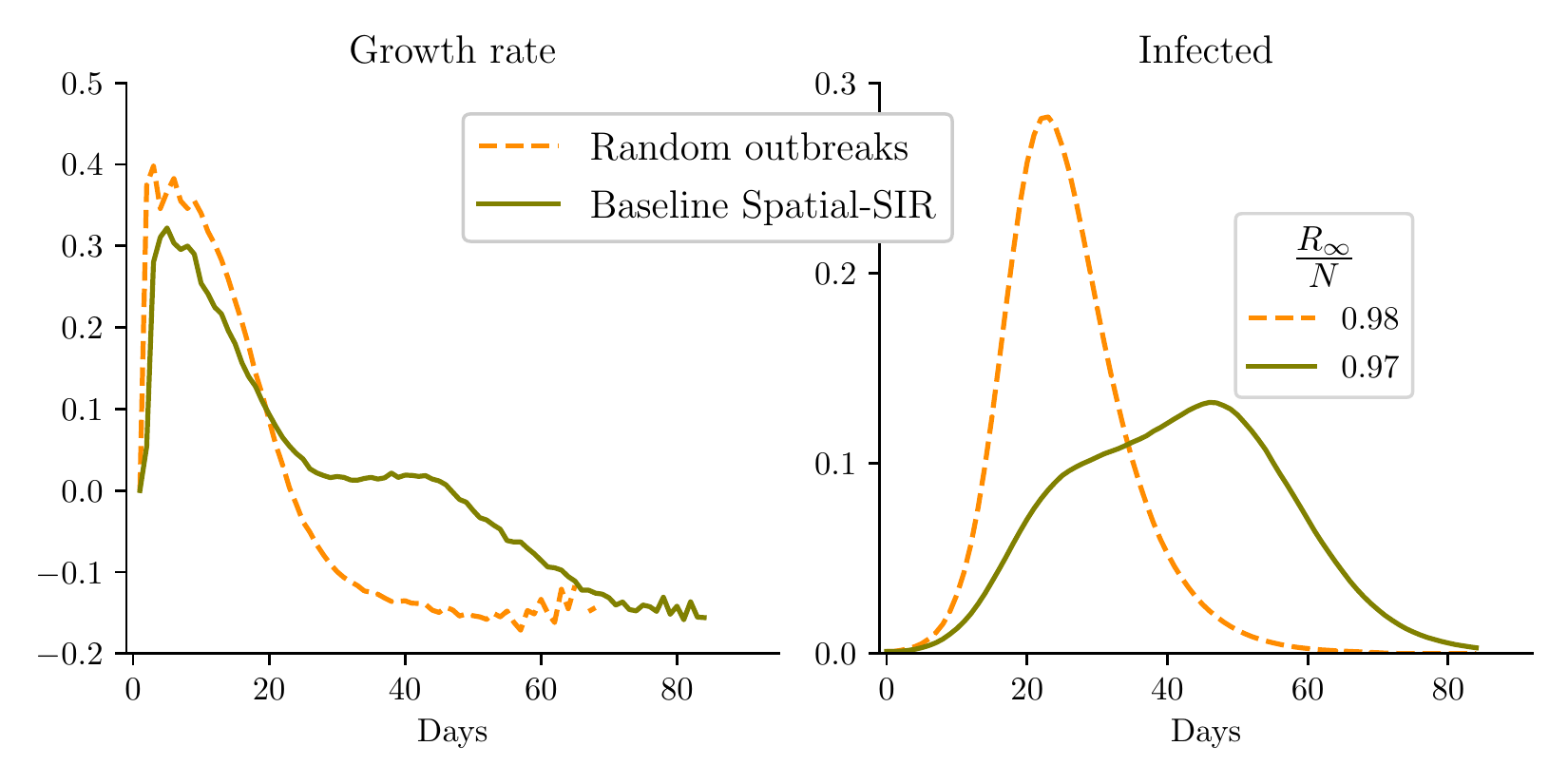}
    \caption*{\normalfont\footnotesize 
    \notegrowth \notelegend\  Solid green lines: spatial-SIR baseline calibration; dashed orange lines: Spatial-SIR with initial outbreaks in random locations}
\end{figure}

To understand the role of the distribution of outbreaks in Spatial-SIR, in Figure \ref{fig:clustercomp}
we compare the progression of the contagion in the baseline city (panel \ref{fig:geo2}, reproduced from Figure \ref{fig:geo}) with the progression in a city in which the infected agents are not placed on an initial cluster but are split in random locations (panel \ref{fig:City-cluster-comparisons}). 
While in the baseline model
contagion is relatively concentrated by day 30, contagion is much more widely spread when the initially infected are randomly located. Figure \ref{fig:cluster-rates} summarizes the infection dynamics in these two simulations: the progression of active cases 
is faster when the initial cluster is randomly located, reaching a higher peak of
 active cases (28\% rather than 13\%) earlier (on day 23 rather than on day 46). However, the fraction of Recovered at the stationary state is about the same, $97\%$).%
\footnote{Please note that in this and other figures, the scales of the x and y axes may differ to improve the visualization. The x-axes are scaled to the number of days it takes for the epidemic to converge to steady-state; the y-axes are scaled to a level slightly greater than the maximum value of the displayed variables}

\subsection{Population size} 

In this section, we study the effects of changing population size $N$ and city area proportionally so as to keep the city
density constant, fixing the number of initial infections, $I_0$, and their spatial distribution (assumed homogeneous to highlight the effects of $N$). 

\begin{figure}[t]
    \caption{Infection dynamics: population size in Spatial-SIR (left panel) and SIR (right panel)}\label{fig:SIR-City-size-comparisons}
    
    \centering{}%
        \includegraphics[width=.85\textwidth]{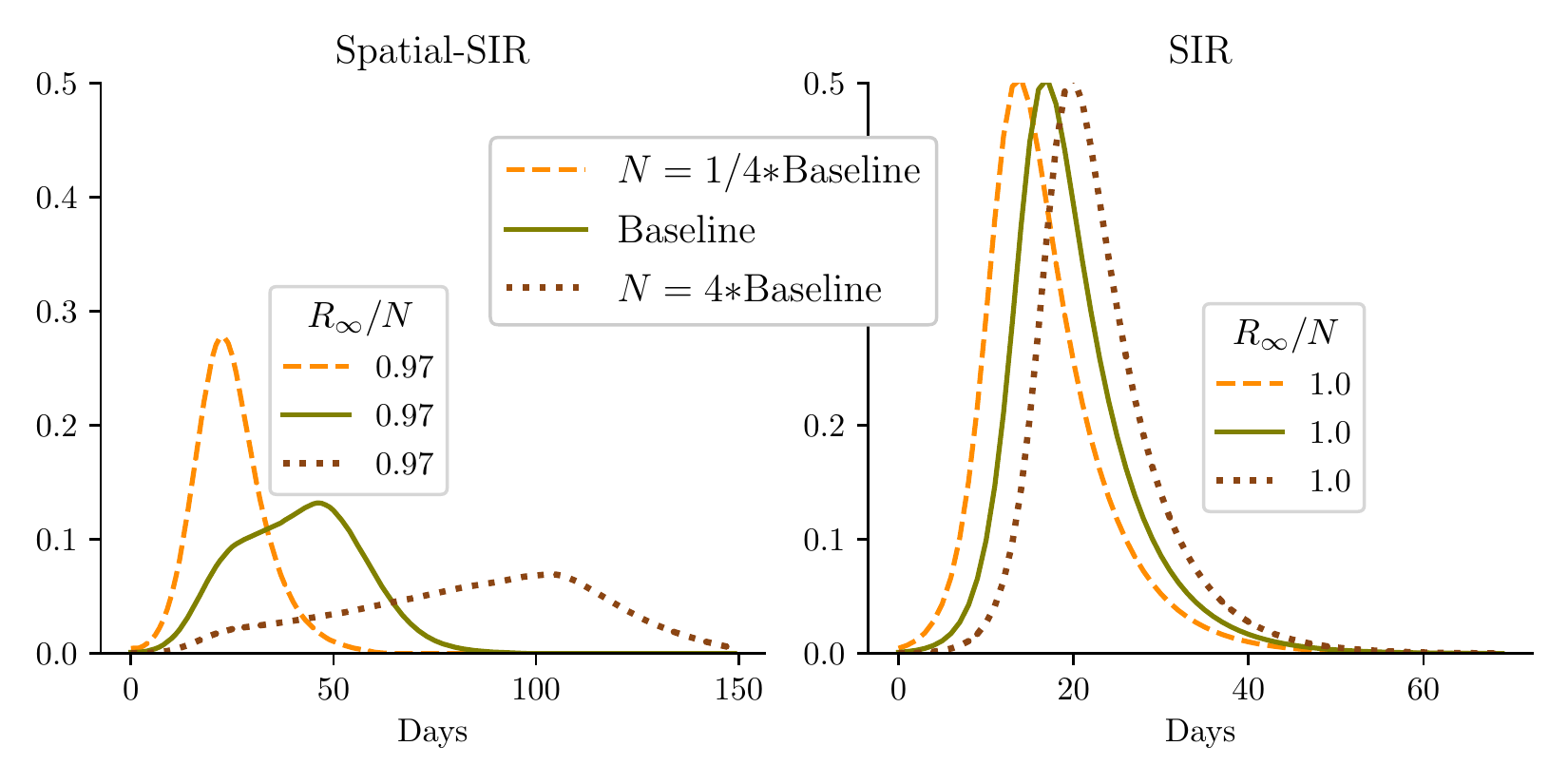}

    \caption*{\normalfont \footnotesize 
    \notemix{. Spatial-SIR (left panel)}{SIR (right panel)} \ Solid green solid lines: spatial-SIR baseline calibration; dashed orange line: one quarter the baseline population; brown dotted lines: four times the baseline population, keeping density constant. \notelegendbis}
\end{figure}

In the SIR model, these changes have no effect on the stationary state (first invariance in Section \ref{invariance}) and a vanishing effect on the transitional dynamics for small enough $I_0$ (second invariance in Section \ref{invariance}).  More generally, increasing size $x$-times  (for a given $I_0$)  increases the peak by only  $-1/ \mathcal{R}_0 \ln x$  percentage points in SIR.   
On the other hand, while population size has no effect on the stationary state of the epidemic in Spatial-SIR as well, it has an important effect on transitory dynamics. 
In Figure \ref{fig:SIR-City-size-comparisons}   we 
report infections, as a fraction of the population, for both models in three cities: the baseline, a city with one fourth, and one with four times the baseline population.  As noted, changing population size does not change the stationary state fraction of infected, approximately equal to
97 percent of the population, independently of city size, in both models. 

With regards to the transitional 
dynamics, their dependence on size is minimal in SIR, hardly visible in fact  in our simulation (see right panel --- more populated cities take longer to reach the peak only because we keep the initial conditions constant). In Spatial-SIR instead, the curve displaying the fraction of active
cases is flatter in larger cities (left panel). More specifically, in Spatial-SIR, with respect to SIR, the same difference in population size reduces the peak to about one quarter (from .28  to .07 active cases). Relatedly, the time (in days) to reach the peak in larger cities goes from 15 to 22 days in SIR, and from 23 to 104 days in Spatial-SIR. Note that the peak is lower in larger cities \emph{as a fraction of the population}, which suggests that resources such as hospital beds, ventilators, etc\ldots, distributed proportionally to population size are less likely to be binding in large cities.

\subsection{City density} 

\begin{figure}[t]
    \caption{Infection dynamics: density, constant $\beta$}\label{times6}
    \centering
            \includegraphics[width=0.85\textwidth]{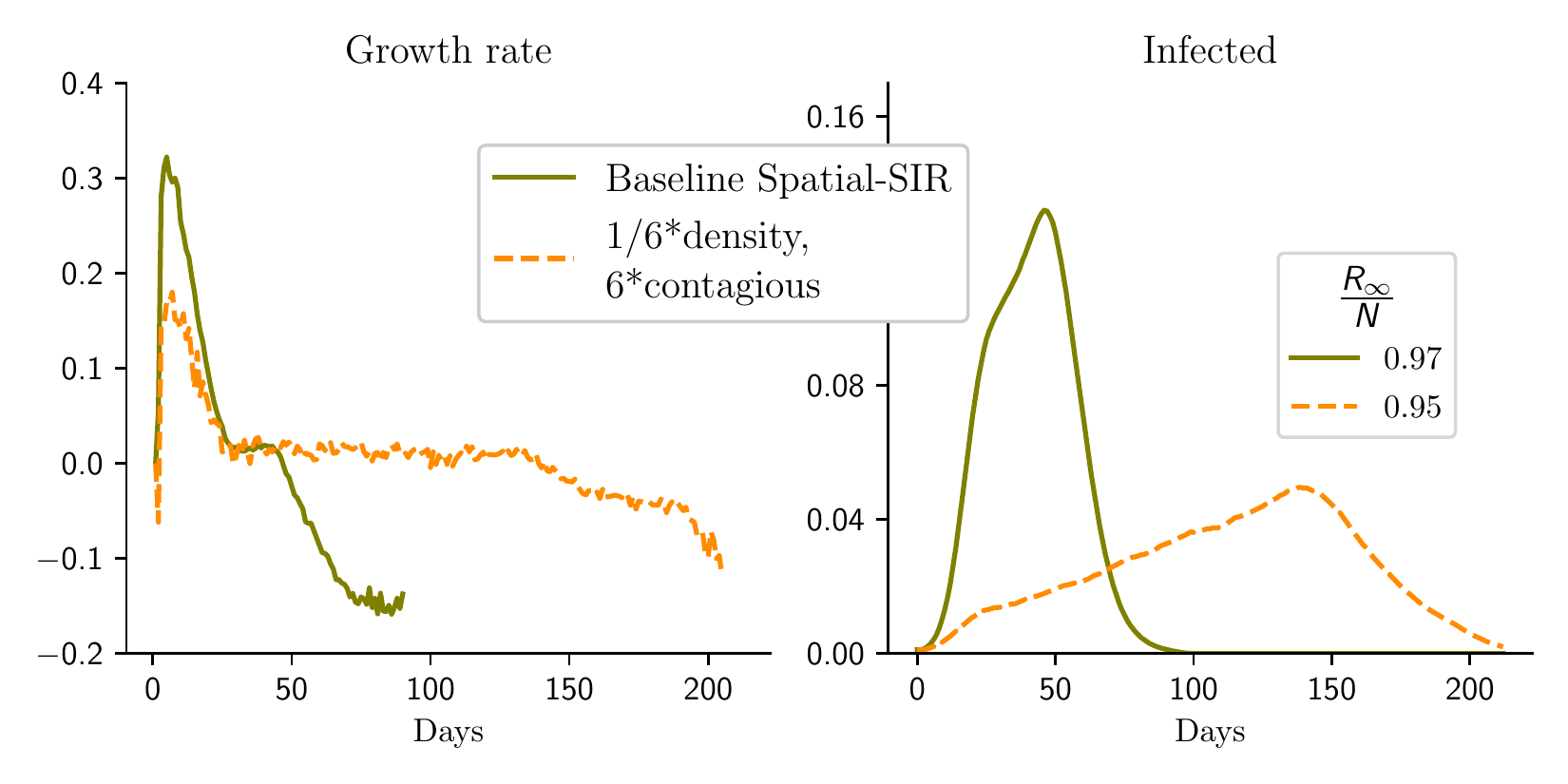}
    
        \caption*{\normalfont \footnotesize 
        \notegrowth \notelegend solid green lines: spatial-SIR baseline calibration; dashed orange lines: baseline Spatial-SIR with one-sixth the density and six times the contagion rate as the baseline model.
        }
        
\end{figure}

In this section, we describe the role of city density on the dynamics of the epidemic. We first show that  city density - which is proportional to contacts - 
plays a distinct role in 
Spatial-SIR from the inverse of the probability of infection, breaking the third SIR  invariance in Section \ref{invariance}.
This is shown in Figure \ref{times6} where the baseline calibrated Spatial-SIR is compared with an environment with six times the probability of infection and 1/6th the density: the effect on the dynamic of 
infection is different both qualitatively and
quantitatively. 

\begin{figure}[t]
    \caption{Infection dynamics: density}\label{fig:3density}
        \includegraphics[width=0.85\textwidth]{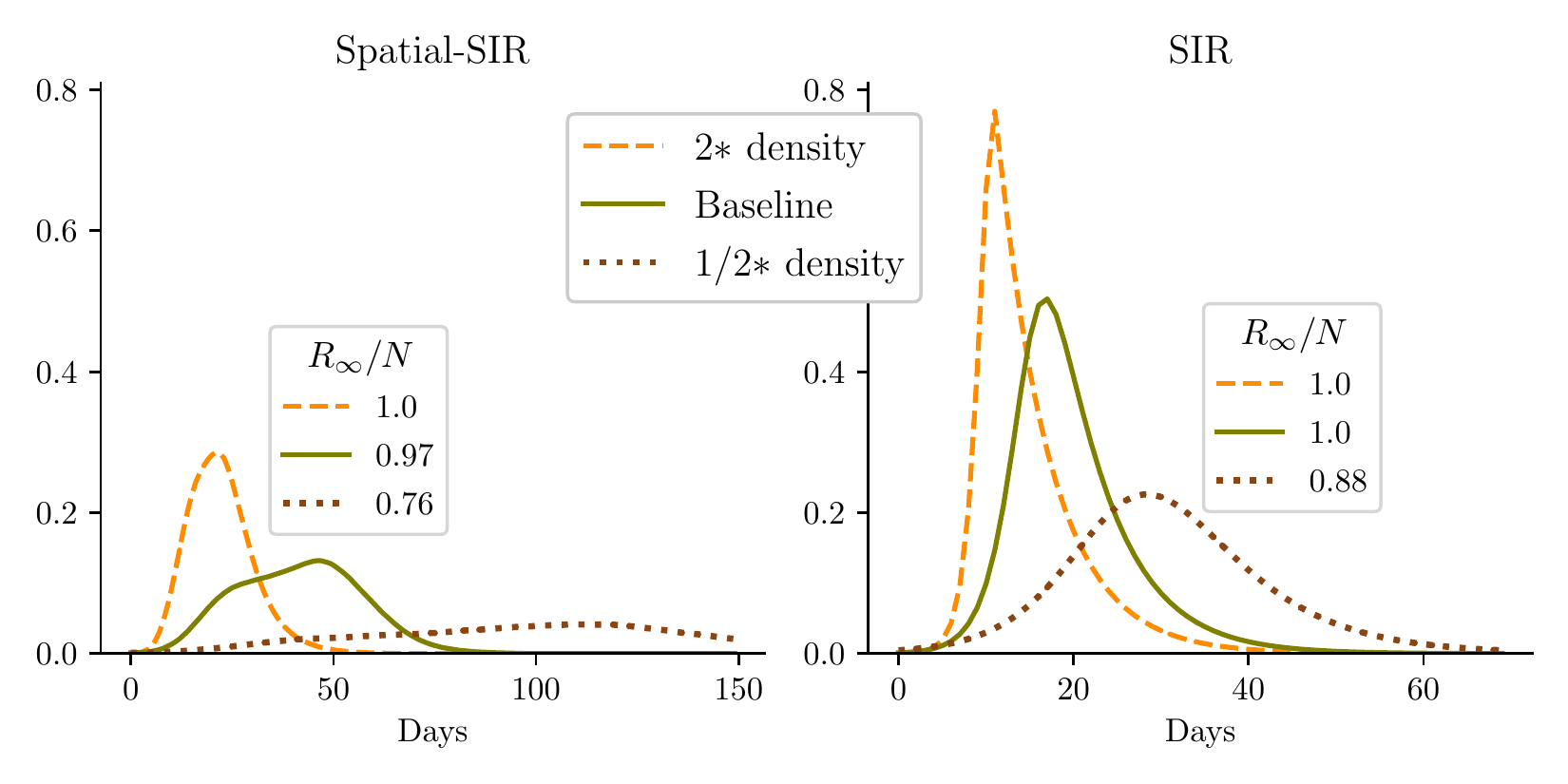}
    \centering
        \caption*{\normalfont \footnotesize 
        \notemix{. Spatial-SIR (left panel) }{and SIR (right panel).}{}{}  Solid green lines: spatial-SIR baseline calibration; dashed orange lines: baseline Spatial-SIR with twice the density; brown dotted lines: baseline Spatial-SIR with one half the density, keeping population size constant. \notelegendbis
        }
      
\end{figure}

Density is a crucial determinant of the dynamics of the epidemic because, together with the contagion
rate, it determines the average number of infections occurring on a given date. Increasing
density while keeping the contagion radius the same increases the number of contacts that each infected individual has on a given day. In fact, in Spatial-SIR, changing city density while keeping the contagion rate and the population size constant has important effects on both the stationary state and the transitional dynamics of the epidemic, as illustrated in Figure \ref{fig:3density} (left panel).  We see that indeed the peak of active cases is increasing in, and very sensitive to, density:   halving density relative to the baseline dramatically flattens the peak of the infection (by more than one-half, after more than twice as many days from the outbreak). 

Comparing the effects of density (through contacts, and in turn through $\beta$) in  SIR and Spatial-SIR reveals some interesting patterns. The effects are qualitatively similar, as can be seen comparing the simulations on the left panel (Spatial-SIR) and on the right panel (SIR) of Figure \ref{fig:3density}. Quantitatively, however, density has much larger effects in Spatial-SIR. In the stationary state limit, the fraction of recovered in the densest economy is 1.13 times larger than in the least dense economy in  SIR and 1.31 times larger in Spatial-SIR. This is the case also with regards to the transition, i) the peak of active cases in the densest economy is 3.45 times larger than in the least dense economy in  SIR and 7.25 times larger (resp. 5.4 times smaller) in Spatial-SIR, and ii) the number of days to the peak on infections in the densest economy is four times smaller than in the least dense economy in  SIR and 5.4 times smaller in Spatial-SIR. 

\begin{figure}[t]
    \caption{Infection dynamics: heterogeneous density}\label{fig:hetdensity}
    \centering
    \begin{subfigure}{.5\textwidth}
        \centering
      \includegraphics[width=.85\textwidth]{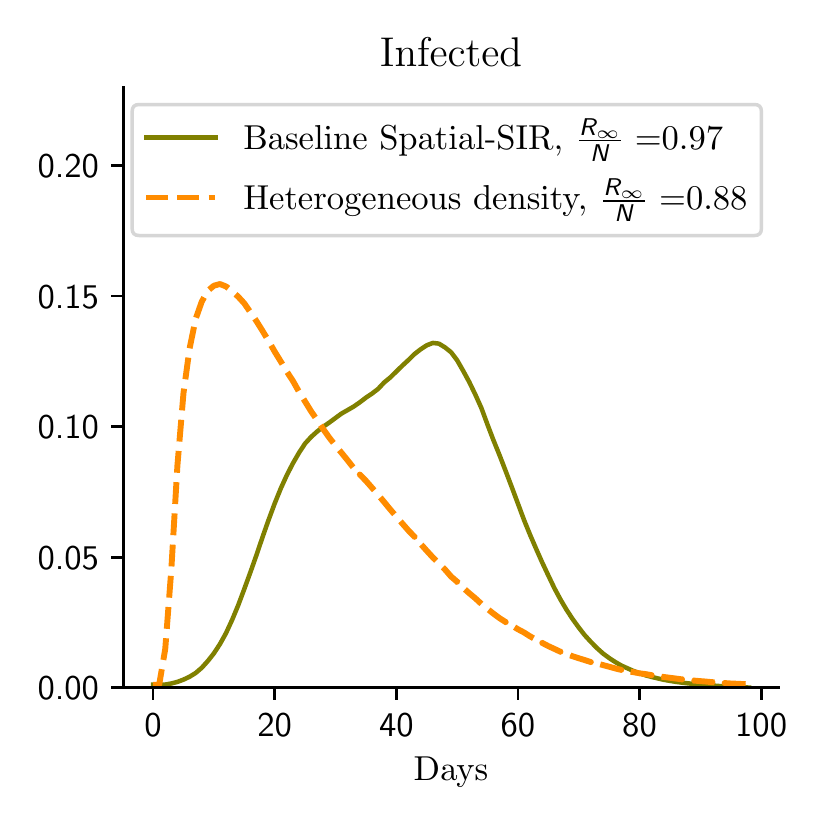} 
    \end{subfigure}%
    \begin{subfigure}{.4\textwidth}
        \centering
 		\includegraphics[width=.85\textwidth]{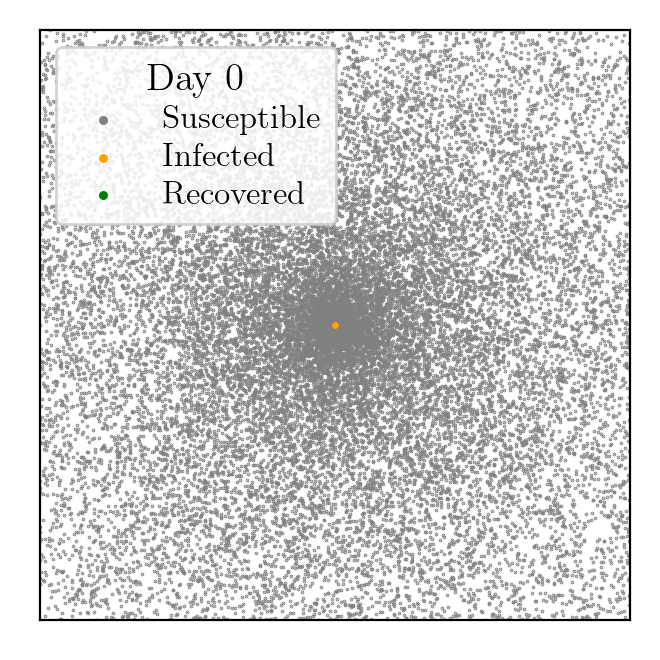}
    \end{subfigure}
    \caption*{\normalfont \footnotesize \notemix{.}{}{}{}Solid green line: spatial-SIR baseline calibration; dashed orange line: Spatial-SIR with heterogeneous density. The 
    right panel 
    illustrates the initial spatial distribution of Susceptibles in the model with heterogeneous density. The legend on the left panel indicates the steady-state fraction of Recovered.}
\end{figure}

We also report, in Figure \ref{fig:hetdensity},  simulations comparing the baseline city (of constant density across space) with an identical city of heterogeneous density, declining from center to periphery (and initial cluster of infection in the center - the right panel illustrates the initial condition, with each grey dot representing a susceptible individual).
While the stationary states of these cities differ minimally, the city with heterogeneous density experiences a smaller peak substantially earlier than the baseline. Namely, heterogeneous density induces a faster-growing epidemic initially, which subsequently slows down,  reaching herd immunity earlier (at about 40\% infected rather than 70\%). 
This example illustrates a more general 
\emph{selection}
mechanism operating when agents are heterogeneous (for example, in age, socio-economic and professional characteristics, preferences for social interactions): those more susceptible to the spread of the infection (in this simulation, those living in denser regions) are selected to achieve herd immunity earlier.\footnote{See \cite{gomes2020individual} and \cite{britton2020mathematical} for  related theoretical analyses.}  

\begin{figure}
    
    \thisfloatpagestyle{empty}
    \caption{Geographic progression of infections and recoveries}\label{fig:nomove}
    
\begin{subfigure}{1.\linewidth}
\makebox[\textwidth][c]{ 
    \begin{tabular}{ccc}
        \includegraphics[width=0.33\linewidth]{figuresdir/nc5-baseline_pos-day10.png} &  
        \includegraphics[width=0.33\linewidth]{figuresdir/nc5-baseline_pos-day20.png} &
        \includegraphics[width=0.33\linewidth]{figuresdir/nc5-baseline_pos-day30.png}
    \end{tabular}
}
    \caption{Baseline model, days 10, 20, 30} 
    \label{fig:geo3}
\end{subfigure} 
\medskip

\begin{subfigure}{1.\linewidth}
    \makebox[\textwidth][c]{ 
    \begin{tabular}{ccc}
    \includegraphics[width=0.33\linewidth]{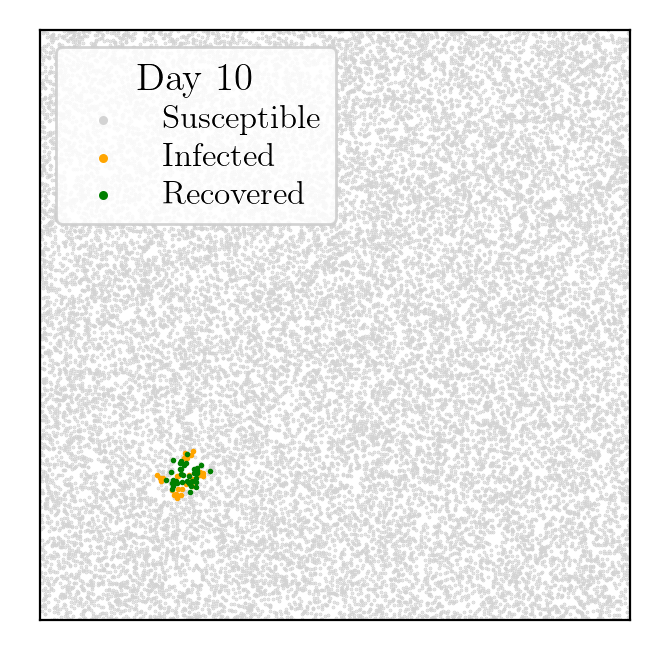} & 
    \includegraphics[width=0.33\linewidth]{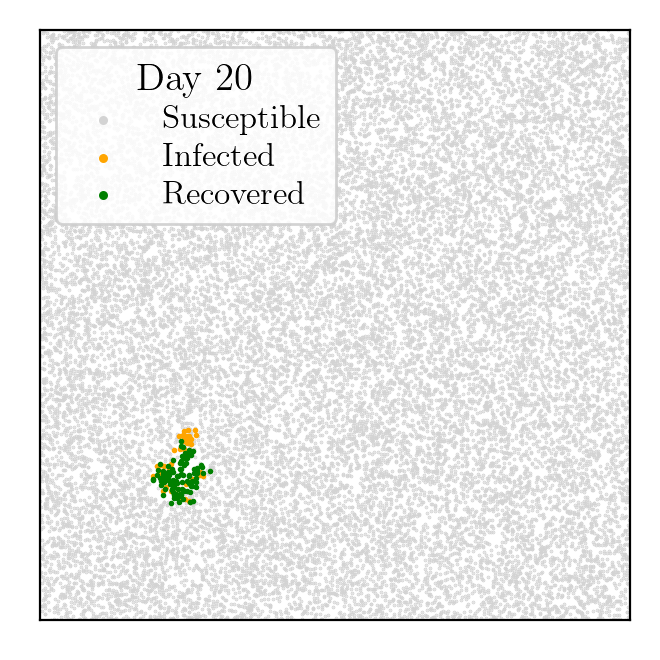} & 
    \includegraphics[width=0.33\linewidth]{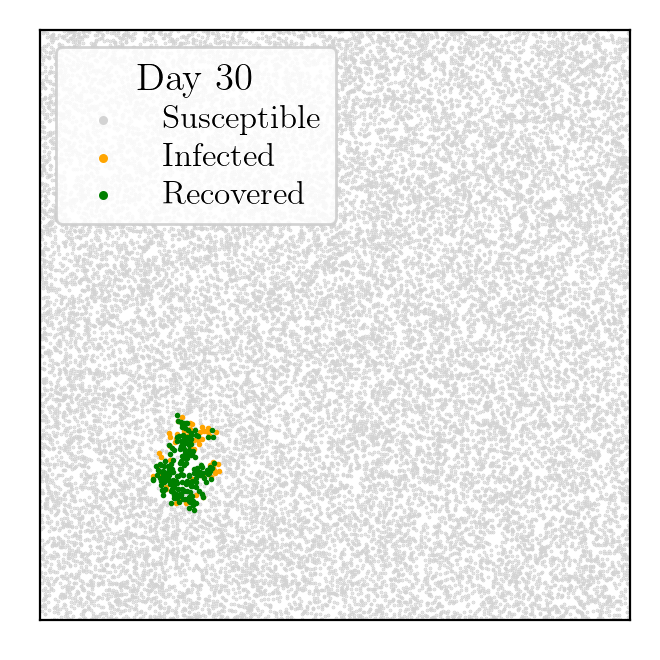}
    \end{tabular}
    }
\end{subfigure} 

\begin{subfigure}{1.\linewidth}
    \makebox[\textwidth][c]{ 
    \begin{tabular}{ccc}   \includegraphics[width=0.33\linewidth]{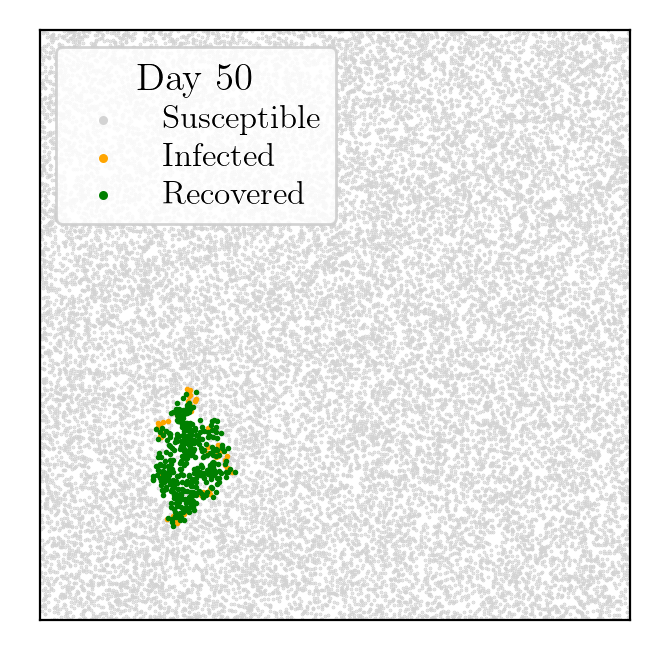} &          \includegraphics[width=0.33\linewidth]{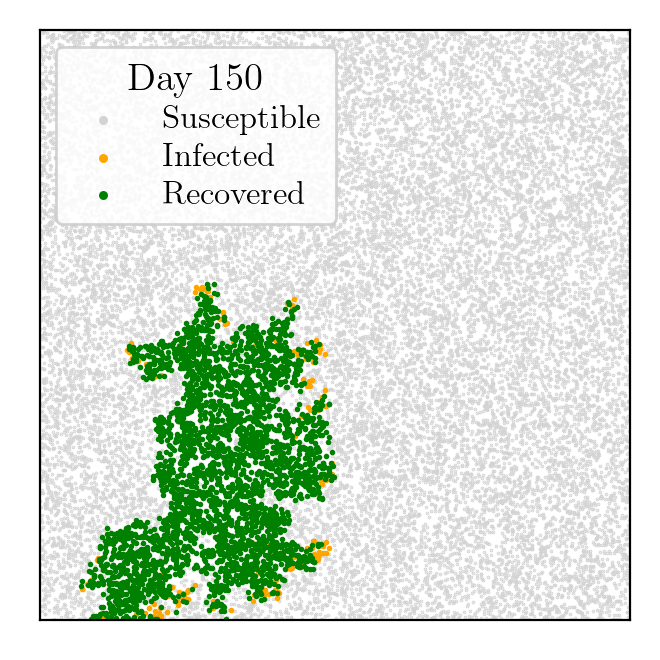} &  		\includegraphics[width=0.33\linewidth]{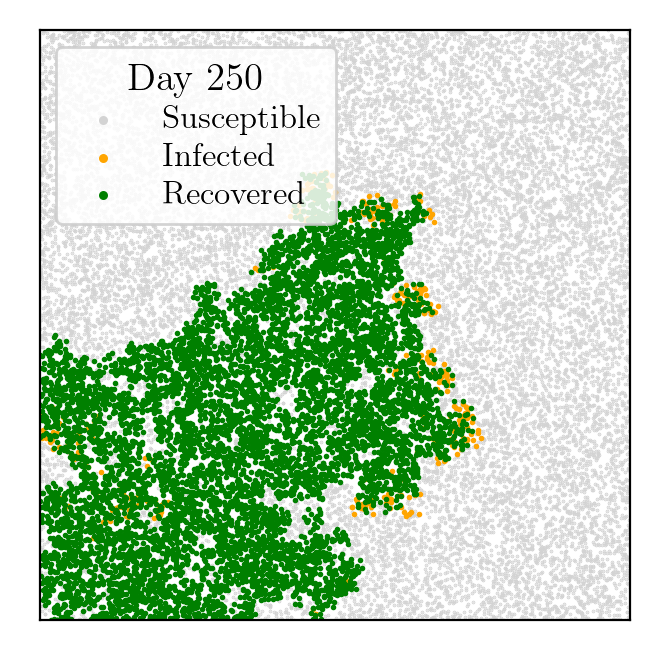}
          \tabularnewline
        \end{tabular}
    }
    \caption{No movement, days 10, 20, 30, 50, 150, 250} \label{fig:nomove-geo}
\end{subfigure} 

   \caption*{\normalfont\footnotesize Note: Each figure displays the position of individuals in the city at a given day since the start of the infection, color-coded by the agent's state}
\end{figure}

\subsection{Movements in the city}

In Spatial-SIR, the parameters controlling the variation in the random movement contribute to explaining the cross-city heterogeneity in the dynamics of the epidemic (these parameters obviously do not appear in SIR). In this section we study the effects of the distance traveled
every day by each agent, $\mu$, on the epidemic dynamics.\footnote{Given our calibrated contagion radius, if all people in the city were uniformly spaced from each other, contagion would not occur. All infections in the baseline model occur because random placement and movement generate clusters of people closer to one another than the infection radius.} Changing these parameters affects the average number of contacts in the city. As we argued, the average number of contacts in the city has an effect that is similar to city density.

\begin{figure}[t]
    \caption{Infection dynamics: movement speed}\label{fig:nomovement-rates}
    \centering

        	\includegraphics[width=0.45\linewidth]{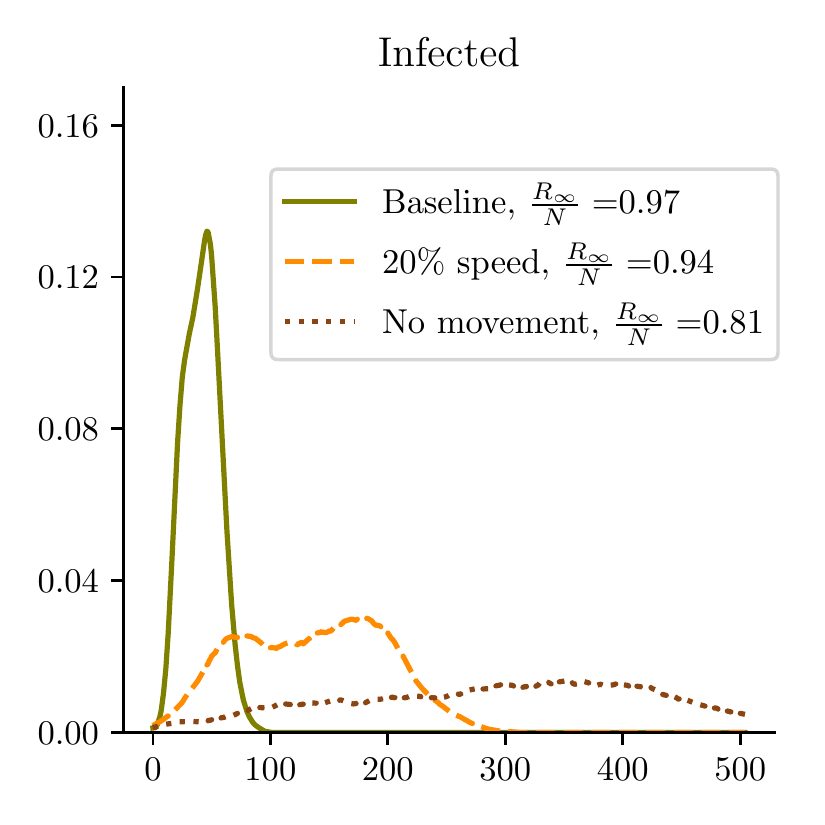} 
    \caption*{\normalfont \footnotesize \notemix{. Solid green line baseline Spatial-SIR;}{ dashed orange line: Spatial-SIR with 20\% speed of movement relative to baseline,}{ and Spatial-SIR with no movement (dotted brown line).}
    The legend indicates the steady-state fraction of Recovered.}

\end{figure}

To provide an intuition of the dependence of the epidemic on the movement speed of agents in the city, in Figure \ref{fig:nomove} we compare the progression of contagion of the baseline model with the same progression in the extreme case when agents do not move.
The infection spreads slowly. As the contagion expands, clusters of 
susceptible (non-infected) people are clearly visible in the rightmost panel as large white spots within the green cloud. This is less likely to occur when people move, which is why 
the speed of movement also affects the steady-state, as illustrated in Figure \ref{fig:nomovement-rates}.

With constant density and people randomly moving around the city, the average number of contacts is 
constant, but local herd immunity plays a fundamental role, and the dynamic of the infection changes with speed. With faster speed, infected people
are more likely to find uninfected locations, making it less likely for people in these
locations to stay immune until the steady-state. 

The speed of people's movement around the city and the number of initial clusters have a 
a very similar effect on outcomes, because if people move very fast, at the beginning of the infection
they generate new clusters quickly.  

\section{Behavioral Spatial-SIR \label{sec:behavioral}} 

As we discussed the introduction, most epidemiological models do not formally account for behavioral responses to the epidemic. In those models, as in the analysis in the previous sections,  the number of daily contacts in the population, $c$, is a constant. 

In this section, we model agents responding to the epidemic by choosing to limit their contacts. Following \cite{keppo2020behavioral}, we introduce a reduced-form behavioral response,  represented by a function $0 \leq \alpha( I_t) \leq 1$, acting as a proportional reduction of the agent's contacts (the population density $d$ multiplied by the contagion area of an infected individual $\Psi$) as a function of  the number of infected in the population:

\begin{equation} c=\alpha( I_t) d \Psi, \; \; \; 
    \alpha(I_{t})=\left\{ 
    \begin{array}{ll}   
        1  & if \; \;  I_{t} \leq \underline{I} \\ 
        \left( \frac{\underline{I}}{I_{t}} \right)^{1-\phi}  & if \; \;  I_{t}>\underline{I}
    \end{array} \right. .  
    \label{Ketal}
\end{equation}

We calibrate the dynamics of the epidemic allowing for the behavioral response (\ref{Ketal}) in  both SIR and Spatial-SIR.\footnote{We calibrated the SIR model as in the simulations in Section
\ref{sec:SIRcomp} for this comparison.} In simulations of the behavioral models  we set $\phi=0.88$ in (\ref{Ketal}) as estimated by
    \cite{keppo2020behavioral} using Swine flu data, and assume people start responding to the spread of the contagion very soon by setting $\underline{I}=0.01$. 
In simulations of the standard SIR model with behavioral
responses, we use the same parameters. We rank individuals by risk aversion and assume that anyone who self-isolates when $I_t=\hat{I_t}$ also self-isolates when $I_t>\hat{I_t}$, inducing persistence in the identity of the individuals who respond. 

\begin{figure}
    \centering
    \caption{Reduction in contacts in behavioral models}
        \label{fig:behavioral}
        \includegraphics[height=.425\textwidth]{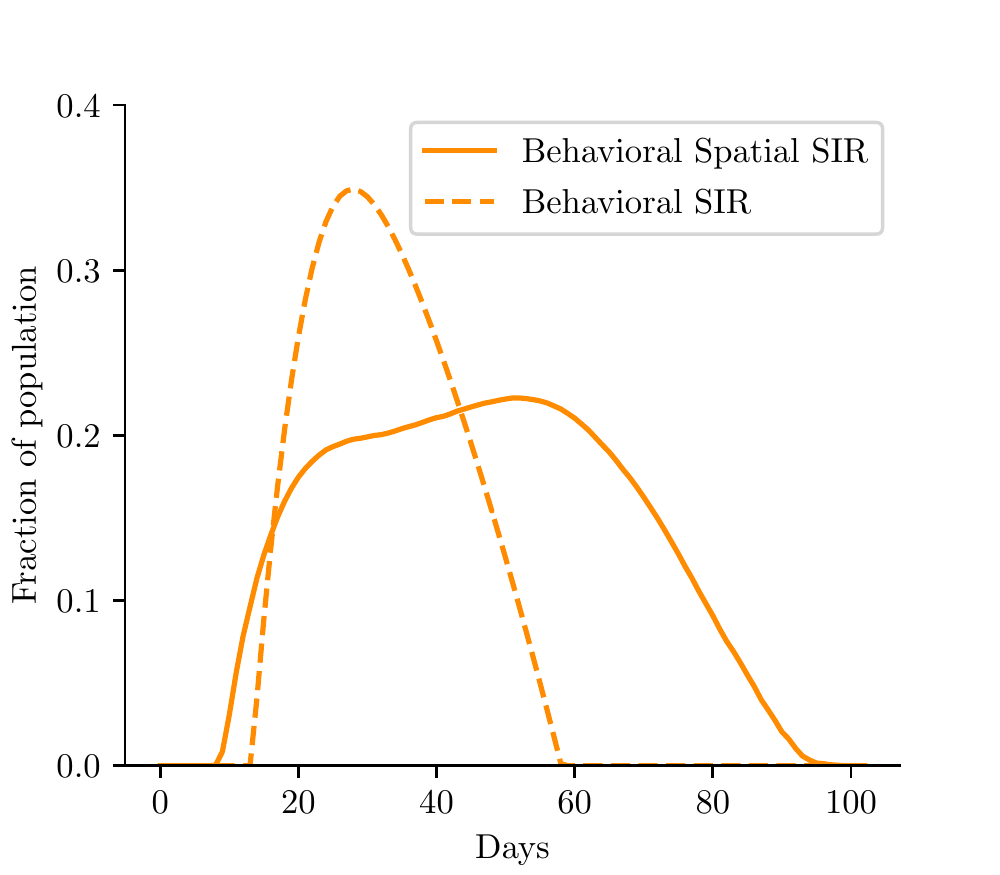}
        \caption*{\normalfont \footnotesize Note: fraction of the population induced to avoid contact due to behavioral responses}
\end{figure}

The simulated   reduction in the number of contacts due to the behavioral response is reported in Figure  \ref{fig:behavioral}. \label{quotebehcomparison} In both the behavioral SIR and the behavioral Spatial-SIR, as the infection spreads, the reduction in contacts due to the behavioral reaction increases. Then, as herd immunity begins and the number of Infected 
declines, the reduction of contacts decreases, and contacts return towards the initial (pre-infection) state. In Spatial-SIR, however, the reduction in contact is (about a third) smaller, but its peak drags for much longer. This is because local herd immunity starts showing its effects earlier, inducing agents to stop reducing contacts earlier, but then builds up more slowly.  

\begin{figure}
    \centering
    \caption{Infection dynamics in SIR and Spatial-SIR, effects of behavioral responses}
    \label{fig:allbeh}
    \begin{subfigure}{\linewidth}
    \makebox[\textwidth][c]{ 
    \includegraphics[height=.375\textwidth]{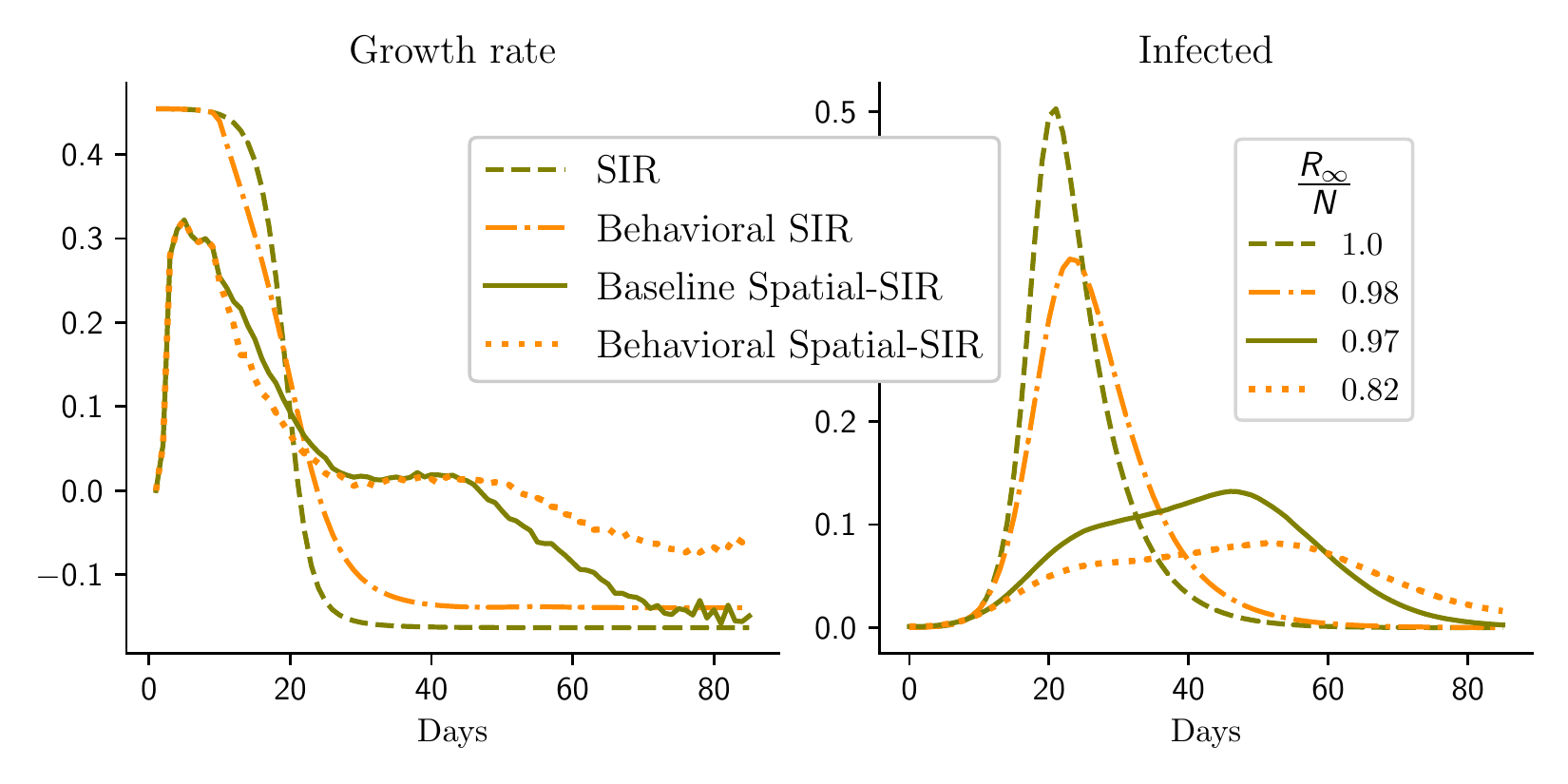}
    }
    \end{subfigure}
    
    \caption*{\normalfont\footnotesize 
    \notegrowth
    \notelegend \ Solid green lines: spatial-SIR baseline calibration; dashed green lines: SIR; dotted orange lines: Behavioral Spatial-SIR; dash-dotted orange lines: Behavioral SIR.}   
\end{figure}

 \label{behintro} Figure   \ref{fig:allbeh} 
 reports simulations of the effects of behavioral responses by comparing models with and without behavior (green and orange lines, respectively). The peak of the growth rate of infections in the behavioral Spatial-SIR is lower than in the behavioral SIR, but it plateaues when declining, after about 25 days. In fact, the growth rate of infections has a much longer declining plateau in the behavioral Spatial-SIR than in the Spatial-SIR, a dramatic display of the effects of the interaction of behavioral and spatial factors in the spread of infection via the dynamics of local herd immunity.\footnote{The interaction of behavioral and spatial factors also modulate the effects of various non-pharmaceutical interventions, e.g., lockdown rules, as studied in \cite{bisinmoro2020}. In particular, substantiating a ``Lucas critique'' argument, the cost of naive discretionary policies ignoring the behavioral responses of agents and firms depend fundamentally on the local herd immunity effects due to the spatial dimension of the dynamics of the epidemic.}  

In both SIR and Spatial-SIR, not surprisingly, the qualitative effects of behavioral response is to reduce the spread of infection, lowering the peak of infected. In Spatial-SIR, however, behavior also has the effect of slowing down the operation of herd immunity. As the number of contacts returns to normal, the behavioral response has lasting effects in the stationary state, reducing total cases more in Spatial-SIR than in SIR. While we do not report simulations to this effect, we notice here the important fact that the behavioral response, when derived from the agents' choice, depends on geographical characteristics $g$ as well, and these affect contacts. We denote the behavioral response then as $\alpha(I_t;g)$. 

 Figure \ref{fig:allbeh} 
highlights   the differential effects of behavioral responses on SIR. 
The behavioral response is not
only much stronger in Spatial-SIR, but qualitatively different when comparing both infection growth rates and the fraction of active cases. The peak of active cases in Spatial-SIR is 1/3 with respect to SIR, but the decline of the infection after the peak is slower. This is the result of the composition of the behavioral response, $\alpha( I_t;g)$, and the local herd immunity factor $ \lambda(H_t;g)$.  The first acts on the number of contacts, while the second acts on the distribution of infected between the contacts.


\subsection{Behavioral Spatial-SIR with local reactions} \label{sec-spSIR-local}
Spatial-SIR allows for a further dimension of the interaction between behavior and local space. Consider the case in which the behavioral reaction of an individual at time $t$ depends only on the fraction of people infected in a circular neighborhood centered at her location and of radius equal to the contagion radius.\footnote{We induce persistence in the identity of those that self-isolate again by ranking individuals by risk aversion. If the formula predicts a reaction by $x_i$\% individuals in a neighborhood centered around person $i$, person $i$ self-isolates if her risk aversion is below the x-th percentile among her neighbors.}

\begin{figure}
    \caption{The effect of local behavioral responses}
    \label{fig:behav-local}
    \centering
    \includegraphics[width=.85\linewidth]{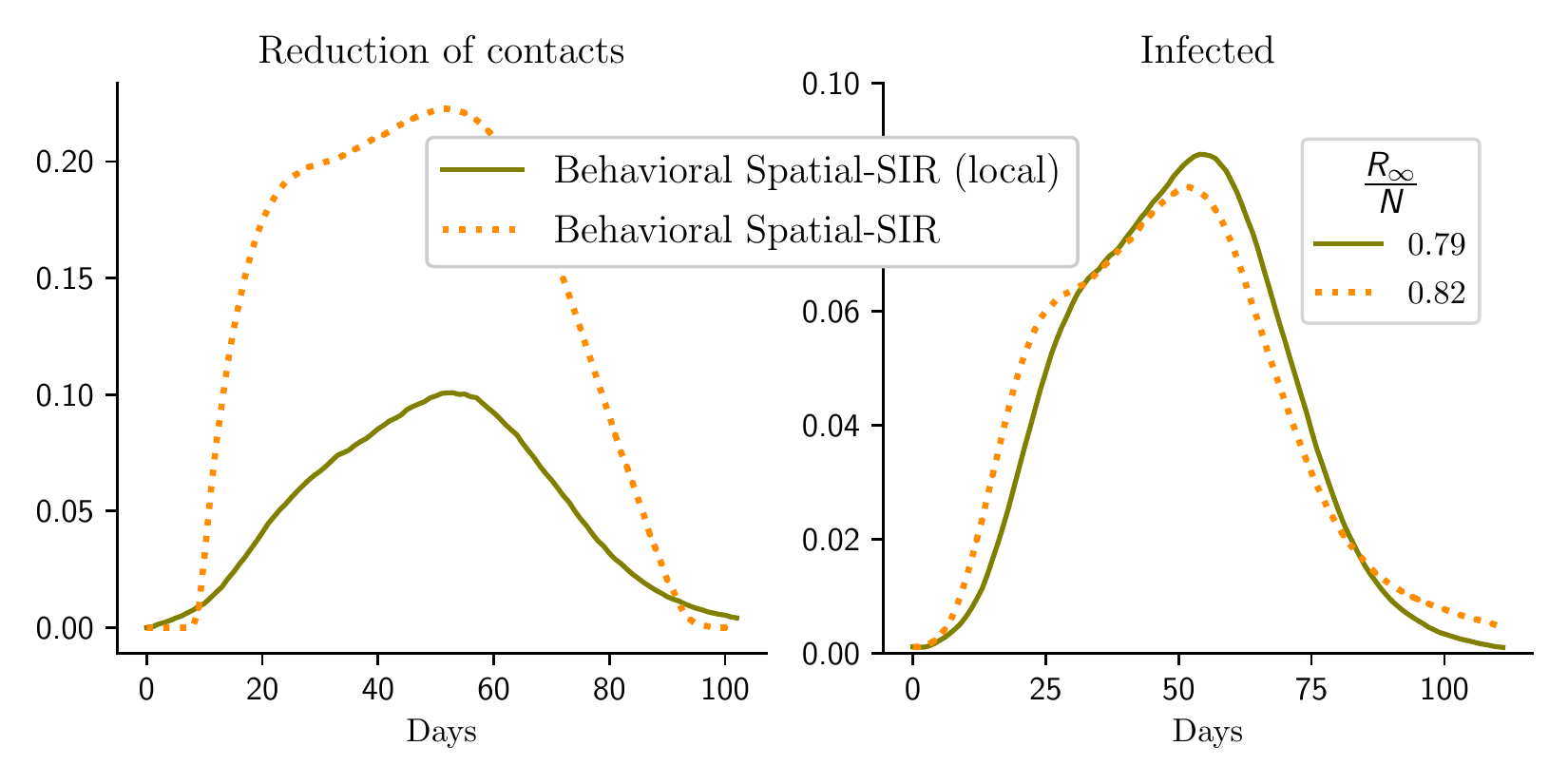}
  \caption*{\normalfont \footnotesize \notemix{. Behavioral Spatial-SIR (colid green line),}{ Behavioral Spatial-SIR with local responses (dotted orange line)}{}{} (right panel).
  The left panel 
    illustrates the reduction in contacts (fraction of population. \notelegend}
\end{figure}

Results are reported in Figure \ref{fig:behav-local}, which reproduces for comparison the results of the behavioral model (the orange dotted line in Figure \ref{fig:allbeh}). 
The heterogeneity of infection rates across neighborhoods has much larger and interesting effects when the behavioral reaction of the agents is local, that is, when it depends on the infection rate in the neighborhood. In this case, as seen in Figure \ref{fig:behav-local}, the higher reduction in contacts during the first and the last days of the epidemic is very short-lived. For most of the epidemic, agents reduce their contacts substantially less (by more than half) with respect to the case in which they react to the infection rate of the whole population because it occurs only in neighborhoods with infections. Interestingly, these composition effects induce relatively small quantitative effects on the population growth rate of infections over time. In other words, local behavioral reactions appear to ``save'' on the reduction in contacts for (almost)-given dynamics of infections. 


\section{Implications for Empirical Analysis}\label{sec:empirical}

We summarize several implications of our analysis to guide empirical research using panel data about the diffusion of an epidemic.\footnote{For illustration, we   concentrate on density as the geographic factor in this analysis. As we showed in the paper, the distribution of outbreaks and the speed of movement of the agents also have systematic effects on the dynamics of an epidemic and are maybe harder to measure. Proxies like airport activity for the number of outbreaks, the distribution of socio-economic characteristics for the distribution of outbreaks, the use of public transportation for the movement of agents, could be fruitfully adopted in both reduced-form and 
structural estimates.}
We first discuss structural estimates of a formal epidemic model. We then  consider  estimates of the causal 
effects of a policy (typically, a Non-Pharmaceutical Intervention 
(NPI), e.g., a lockdown), which in many applications adopt a
Difference in Differences (DiD) design.  

\subsection{Structural Estimates} 

Consider panel data on the dynamics of an infection across different geographic units $i$. The econometrician observes the geographic characteristics $g_i=[I_{i,0}, N_i, d_i, \mu_i]$ of each city $i$
, and data on the dynamics of the infection,   $I_{i,t}, R_{i,t}$ (hence $S_{i,t}$).

\subsubsection*{Cross-city restrictions in the standard SIR.} 

With these data, consider estimating a SIR model without behavioral
effects as in standard epidemiological studies.  Consider the following specification of the growth rate of infections, derived directly from (\ref{idot}): 
\begin{eqnarray} 
    \frac{\Delta I_{i,t}}{I_{i,t}} &=& \beta_{i,t} \frac{S_{i,t}}{N_i} -\rho,   \label{bbetait0} \\
    \text{where } \beta_{i,t}&=&\beta_i    =  \pi c_i, \; \;  c_i=d_i \Psi .   
    \label{bbetait}
\end{eqnarray}  
 
This formulation highlights how model restrictions could be exploited for empirical analysis. Indeed,  Equations 
(\ref{bbetait0}-\ref{bbetait}) impose falsifiable cross-city restrictions, e.g., along the lines of the invariances we identified in Section \ref{invariance}. Several other empirically testable restrictions can be  obtained directly from the closed-form representation of the dynamics; e.g.,  the growth rate of the fraction of infected in a city is, \emph{ceteris paribus}, proportional to the density of the city. 

\subsubsection*{Identifying  local spatial factors.} If the data are generated by a Spatial-SIR, however, accounting for spatial structure becomes important to capture matching frictions through local social interactions, as shown in Section \ref{sec:spatialSIR}. The dynamics of the  infection 
takes the form 
 \begin{eqnarray}
    \frac{\Delta I_{i,t}}{I_{i,t}}  &= &\beta_{i,t}  \lambda (H_t;g_i)   S_{i,t} 
         -\rho, \label{ISS}   \\
 \text {where } \beta_{i,t}&= & \beta_i \text{ as in (\ref{bbetait})}. \nonumber
         \end{eqnarray}  
         
The main driver of the differential effects in Spatial-SIR is local herd immunity. Geographic characteristics $g_i$ mediate the relationship between parameters and model outcomes without a parametric expression for the function $\lambda$, nor for the transition matrix of the stochastic process $H_t$, making it challenging to identify the effects of geography from infection strength separately.
However, one can use the structure of the model
to match data with model predictions using simulation methods. 
Alternatively,
one could use simulations to estimate $\lambda(H_t; g_i)$ which can be used as a correction to the (much faster to simulate) dynamics
of the SIR model, to estimate (\ref{ISS}-\ref{bbetait}).\footnote{Note that, besides density, various distributions of outbreaks can be easily included in the estimation of a Spatial-SIR (but not in the estimation of the SIR).} 

\subsubsection*{Identifying behavioral responses.} 
Finally, accounting for agents' behavioral response to the epidemic in the data  is generally of first order importance, as we discussed in Section \ref{sec:behavioral}. In this case, the formal representation of the dynamics of the infection, Equation (\ref{bbetait0}) in SIR and (\ref{ISS}) in Spatial-SIR, are unchanged, but the number of contacts is endogenous and   (\ref{bbetait}) is modified to  
\begin{equation} 
    \beta_{i,t}    =  \pi c_{i,t}, \; \;  c_{i,t}= \alpha (I_t;g_i) d_i \Psi. \label{aa} 
\end{equation}
 This amplifies the 
issues we highlighted so far, requiring a new identification strategy, notably, to handle the dependence of $\beta_{i,t}$ on $t$.
In SIR the parameters predict the infection dynamics precisely. For example, there is a one-to-one correspondence between initial infection growth rates and the peak. 
Deviations from such dynamics can non-parametrically identify $\pi$ from 
$\alpha (I_t;g_i)$. Parametric identification
can be achieved by assuming a functional form for 
$\alpha (I_t;g_i)$ along the lines of (\ref{Ketal}).
In Spatial-SIR the full specification is (\ref{ISS}-\ref{aa}). Identification in this case can
rely on simulation methods, as needed for the identification of spatial factors.\footnote{\cite{fernandez2020estimating} adopt simulation methods to estimate parameters separately for each location without imposing geographic restrictions.}
 Evidence of agents' movements, using ``Big-Data''
from Google, Safegraph, and Cuebig
could also provide useful empirical strategies for identifying behavioral responses from infection 
dynamics exploiting restrictions imposed by Spatial-SIR.%
\footnote{See, e.g., \cite{farboodi2020}.}

\begin{figure}[t]
    \caption{Estimates of $\beta$ from simulated data}
    \label{fig:est_densitybetas}
    \centering

    \includegraphics[width=.425\linewidth]{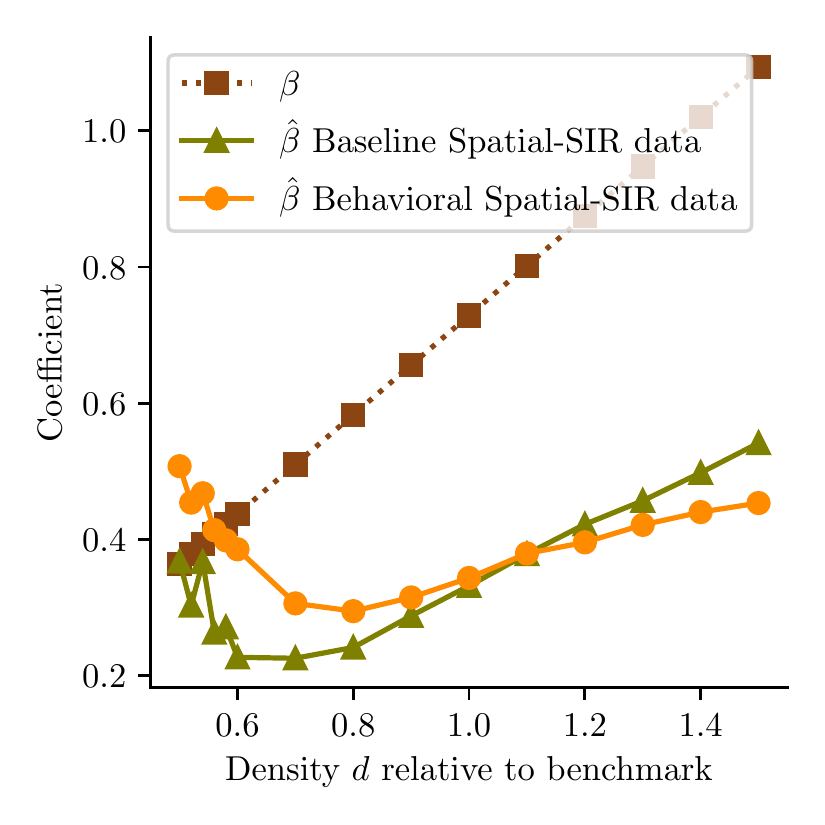}
    \caption*{\normalfont \footnotesize Note: Estimates of $\beta$ using simulated data from models with baseline parameters, except for different levels of population density from 50\% to 150\% of the baseline density, obtained by varying varying the city size. 
    Green line: Spatial-SIR, Orange line: Behavioral Spatial-SIR, The brown line reports the corresponding SIR infection parameter $\beta=c\pi d$.}
\end{figure}

To examine the importance of local spatial factors and behavioral responses in  empirical research exploiting structural models of epidemic, we present some simulation experiments. In particular, we report on the biases occurring when 
estimating (\ref{bbetait0}) on data generated by a Spatial-SIR, with and without behavioral responses of agents to the epidemic. 

Figure \ref{fig:est_densitybetas} reports the values of the estimates of the coefficients on variable $S_{i,t}/N_i$ for cities with different densities from 0.5 to 1.5 relative to the baseline specification density (we do not report confidence intervals because they are very small). We simulate each city 5 times and stop the simulations after the first 80 days, for a total of 400 observations per density level. Interestingly, the mis-specification bias in the estimation is small in cities with small density (where spatial and behavioral factors are less relevant);  but very large for larger densities:  $\hat{\beta}$ is about half the actual $\beta$ of the data generating process for cities with $50\%$ larger  density than the baseline. 

\begin{figure}[th!]
    \caption{Infected at $t$, difference between models without and with lockdown}
    \label{fig:est_betapolicies}
    \centering
    \includegraphics[width=\linewidth]{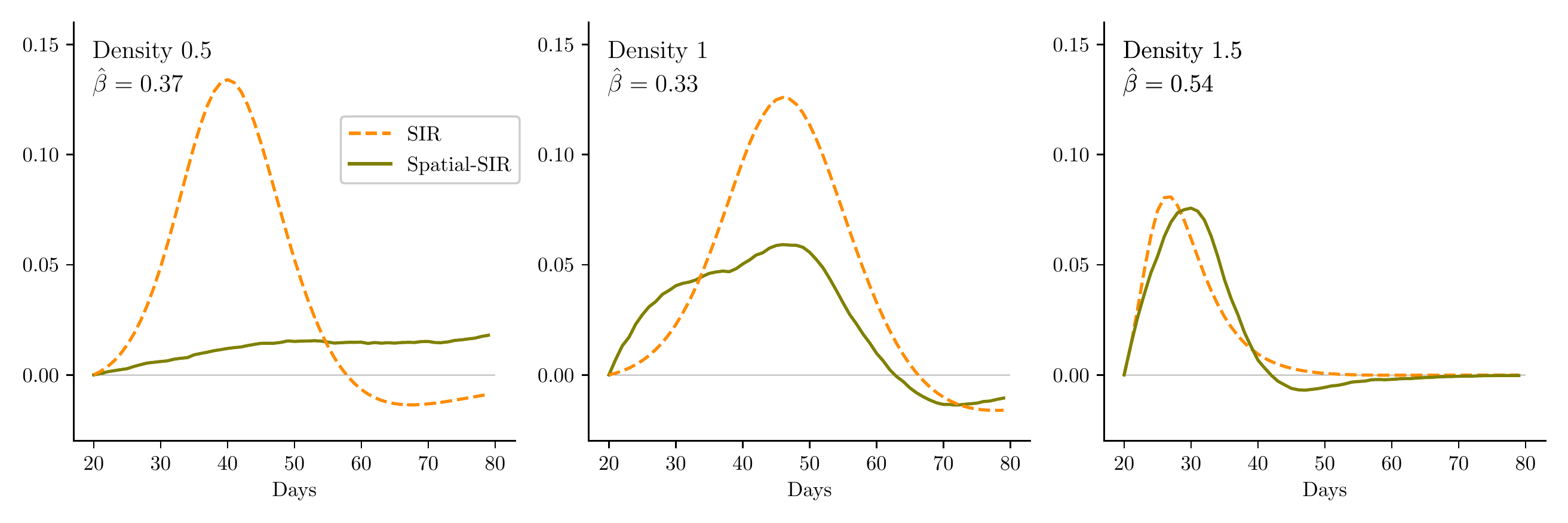}
    \includegraphics[width=\linewidth]{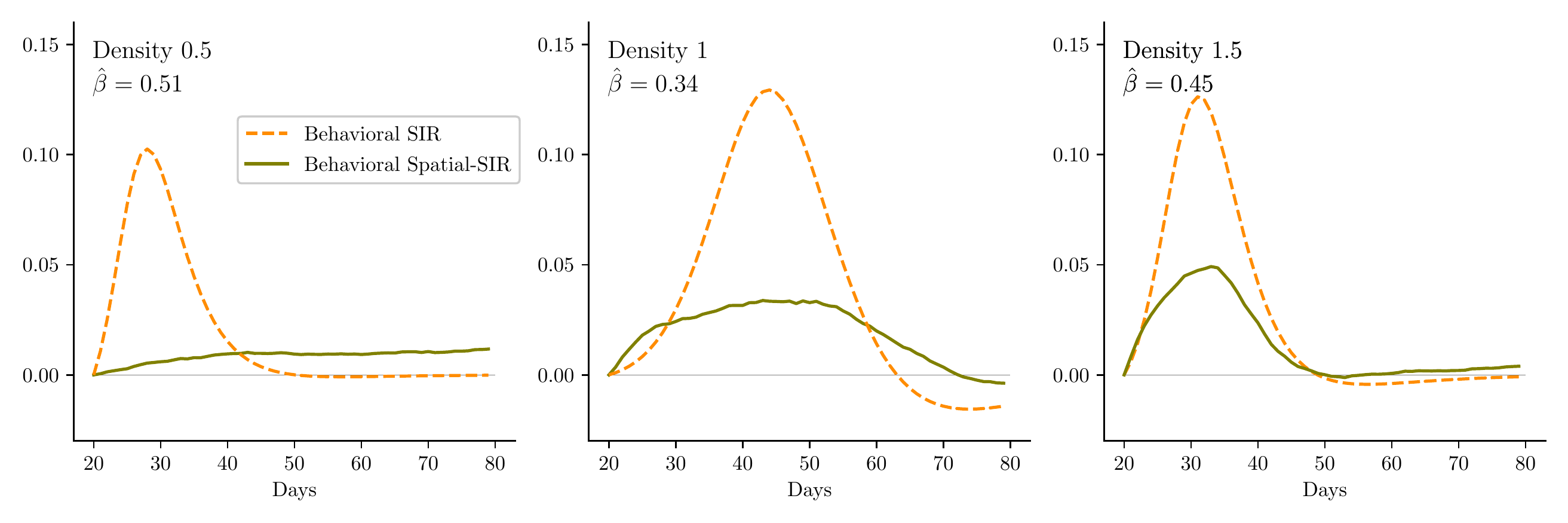}
    
     \caption*{\normalfont \footnotesize \emph{Note:} Difference between infected at date $t$ without and with a 25\% lockdown policy applied at date $t=20$ (fraction of total population). The SIR models are simulated with infection parameter $\hat{\beta}$ estimated using the data simulated from the Spatial-SIR, as reported in Figure \ref{fig:est_densitybetas}. The middle panels are computed with baseline parameters. The left and right-panels Spatial-SIR are computed by varying the citysize to obtain 50\% and 150\% population density, respectively. }
\end{figure}

In Figure \ref{fig:est_betapolicies} we illustrate the effects of this mis-specification bias in predicting the dynamics of active cases after a policy intervention (a $25\%$ lockdown at day 20). The solid green lines illustrate the dynamics of the \emph{difference} in active cases between the Spatial-SIR model without and with the policy. The dashed yellow lines report the same dynamics in the SIR model, at the estimated values $\hat{\beta}$. We report results for only three density levels, and for the models without (three top panels) and with behavioral responses (bottom panels). The implied errors in the effects of the policy are large and have a complex dynamic structure, to the point of changing in sign in some instances.
Note that when density is high (rightmost panels), $\beta$ is high, therefore both SIR and Spatial-SIR have relatively fast dynamics (not visible in this figure). In this case, the effect of a lockdown is quantitatively similar, at least in the model without behavior. Models with relatively low density (smaller $\beta$), despite being more precisely estimated (see Figure \ref{fig:est_densitybetas}), generate relatively fast dynamics in SIR, but not in Spatial-SIR, implying a substantial difference in the effect of the policy, as the leftmost panels illustrate. 
Therefore, we conclude that it is hard to capture the dynamics implied by the spatial structure and the induced local herd behavior with a SIR model with a general reduction in the accessibility of the susceptible population (a lower $\beta$).

\subsection{Causal 
Effects of Policy Interventions}

We simulate the policy intervention we just introduced (a $25\%$ lockdown) as a treatment introduced in some cities at day $20$,  but possibly not in others or introduced at different times in different cities. 

\subsubsection*{Identifying the time-varying effect of geography in DiD studies of policy interventions.} \label{didstudies} 
Let  $\text{Treated}_{i,t}$ take value $1$ if city $i$ is treated by the policy at time $t$. 
Denoting with $Y_{i,t}$ the variable of interest, 
the effects of  $\text{Treated}_{i,t}$ can be evaluated by means of  a DiD design: 
\begin{equation} 
    Y_{i,t}= \nu + \eta_{i}   + \gamma_t +  \delta \text{Treated}_{i,t} +\lambda X_{i,t} \label{ns} 
\end{equation} 
where   $ \nu, \eta_i, \gamma_t$ are time and location effects and $X_{i,t}$ are  controls.
Our analysis suggests that the validity of this research design depends on both the variable of interest, $Y_{i,t}$, and the modeling framework. Consider attempting to estimate the effect of policies on the number of contacts, $Y_{i,t}=c_{i,t}$, for example. In the standard SIR model, this variable is proportional to $\beta_{i}$ and depends on time $t$ only through the treatment. In this case therefore specification (\ref{ns}) can flexibly capture variation in $\beta_i$ by location. 

Importantly, however, this specification fails to capture the dynamics of contacts in a SIR model with behavioral responses by agents. In this case in fact, contacts $c_{i,t}$ is proportional to $\alpha(I_{i,t}; g_i)$, as in equation (\ref{aa}), and hence it is not separable in $i$ and $t$, as required by the additive
form $\eta_i+\gamma_t$ in (\ref{ns}). Policy and agent behavior have separate effects on the dynamic of the epidemic both because behavioral responses have
time-varying effects and because their effects 
interact with 
the effects of geography (a point
generally disregarded in the few studies that try to account for behavioral responses).\footnote{When the data is treated by policy, special care must be used because 
$\alpha (I_t;g_i)$ is also not invariant to policy by a Lucas critique argument, even in the absence of geographical factors; see \cite{bisinmoro2020} for an analysis of this issue.} 

Our analysis shows that including a spatial dimension, as in the Spatial-SIR model, the vector of geographic factors  $g_i$  affects outcomes differently over time, introducing a time-varying heterogeneity that is not fully accounted for by time and location fixed effects as in (\ref{ns}), even after including a vector of geographic factors  $g_i$  among the set of regressors, and even if interacted with time. Furthermore, the direct effects of the treatment also depend on geographic characteristics: a lockdown, for instance, acts as a reduction of density and affects local herd immunity differently depending on the initial density.

Finally, identifying the effects of policy intervention in reduced-form econometric models are even more severe when the variable of interest $Y_{i,t}$ is the growth rate of infections, $\Delta I_{i,t}/I_{i,t}$. In this case, in fact, 
 the structure imposed by SIR, as in (\ref{bbetait0}), generates time-varying heterogeneity not fully captured by  time and location fixed effects even without behavioral responses or spatial dimensions and local herd immunity.

\begin{figure}[t]
    \caption{True and predicted outcomes after a lockdown: time-heterogeneity only}
    \label{fig:prediction_nobias}
    \centering
    \includegraphics[width=\linewidth]{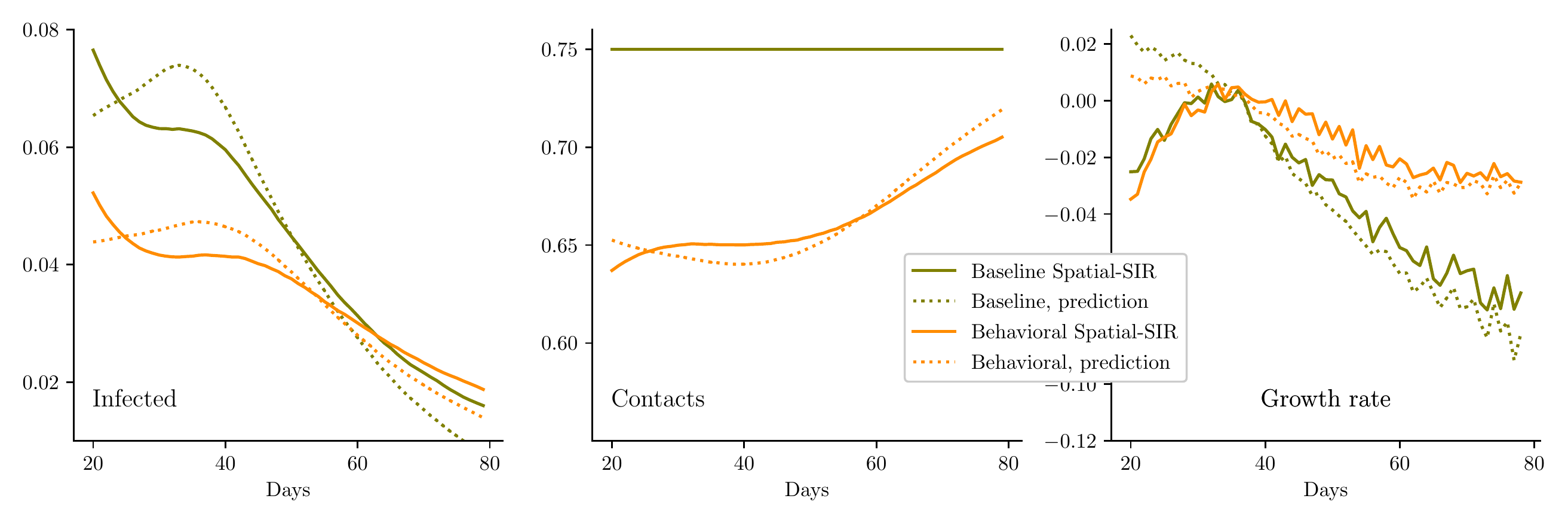}
    \caption*{\normalfont \footnotesize \emph{Note:} Infected (left panel), contacts (middle panel), and growth rate of infected (right panel) at date $t$ (fraction of population in Spatial-SIR (green lines), and Behavioral Spatial-SIR (orange lines). The colid lines simulate the outcomes after a 25\% lockdown imposed at $t=20$. The dotted lines simulate the average by date $t$ of the predicted value of $Y_t$ after estimating (\ref{ns}) using the same data simulated from the Spatial-SIR models of Figure \ref{fig:est_densitybetas}. The dataset includes 80 dates for each of 10 cities (5 treated and 5 untreated) per 11 density levels, for a total of 4400 observations.  }
\end{figure}

To better examine quantitatively the biases induced by estimating the ``two-way'' dif\-fer\-ence-in-dif\-fer\-ences equation (\ref{ns}) we produce several empirical experiments on simulated data. We estimate (\ref{ns}) using the same data we simulated to generate Figures \ref{fig:est_densitybetas} and  \ref{fig:est_betapolicies}; that is, several cities with densities from 0.5 to to 1.5, with and without a 25\% lockdown policy applied at date $t=20$. First of all we study the case in which the data contain, for each density, both cities with and without lockdown; that is, treated and untreated. In this case, therefore, spatial heterogeneity is fully controlled for in Equation \ref{ns}.   Figure \ref{fig:prediction_nobias} reports, for three different outcome variables (active cases, contacts, growth rate of active cases, respectively in the left, center, and right panel) the dynamics after the lockdown is applied, and its predicted values using the estimated coefficients of Equation \ref{ns}, averaged over time. 

\begin{figure}[t]
    \caption{True and predicted outcomes after a lockdown: both time and spatial heterogeneity}
    \label{fig:prediction_withbias}
    \centering
    \includegraphics[width=\linewidth]{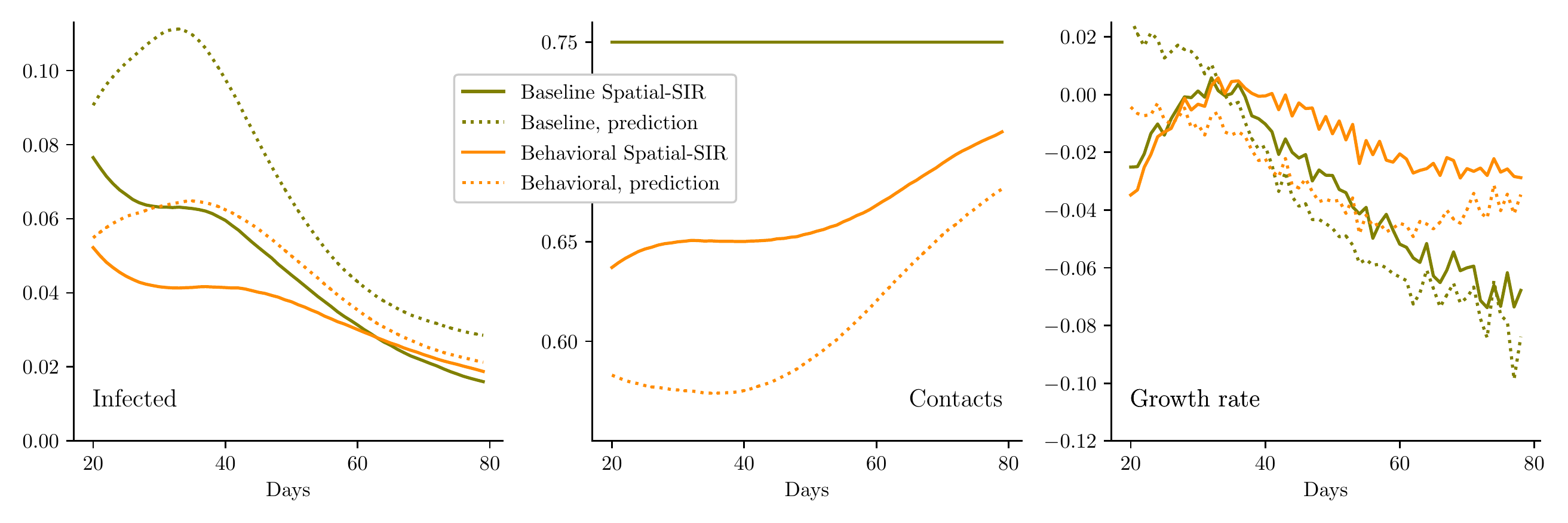}
        \caption*{\normalfont \footnotesize \emph{Note:} Infected (left panel), contacts (middle panel), and growth rate of infected (right panel) at date $t$ (fraction of population in Spatial-SIR (green lines), and Behavioral Spatial-SIR (orange lines). The colid lines simulate the outcomes after a 25\% lockdown imposed at $t=20$. The dotted lines simulate the average by date $t$ of the predicted value of $Y_t$ after estimating (\ref{ns}) using the data simulated from the Spatial-SIR models of Figure \ref{fig:est_densitybetas}, but excluding data from cities treated by the policy if their density is equal or greater than 1, and excluding data from cities not treated by the policy if their density is less than 1. The resulting dataset includes 80 dates for each of 5 cities per 11 density levels, 25 treated and 30 untreated, for a total of 4400 observations.  }
\end{figure}

In this case, effectively controlling for spatial heterogeneity, the estimated equation i) predicts contacts correctly when the data does not contain a behavioral component; ii) but does  not predict contacts with behavior correctly; iii) nor it predicts correctly active cases or their growth rate (with or without behavior).  \label{ate} By construction, Equation \ref{ns} captures the average treatment effects for all variables correctly (the average values of the colid and dotted lines are the same), but fails to capture the  time-heterogeneity of the effects of the lockdown. Figure \ref{fig:prediction_withbias} reports the same simulation experiment, but with estimates computed from data that do not allow to fully control for spatial heterogeneity; that is, data do not contain both treated and untreated cities for each density level.    Specifically, we computed an extreme case assuming that the econometricians observe only treated cities if their density is below 1, and only untreated cities if their density is equal or above 1. In this case, the figure illustrates that the estimated value of the treatment parameter fails to identify even the average treatment effect and prediction errors are quantitatively much larger. This is particularly the case when predicting active cases (left panel): after 35 days, the over-prediction of the estimated equation with respect to the simulated data is of the order of $50\%$ (both with and without behavioral responses). But even contacts (central panel) are quite imprecisely predicted when the data are generated by the model with behavioral responses of agents (as noted, without behavior, Equation  \ref{ns} predicts contacts precisely). 

To better understand the source of prediction error in these exercises, let's focus on the prediction of active cases, that is, on the left panels in figures \ref{fig:prediction_nobias} and  \ref{fig:prediction_withbias}. At the time the policy is applied, $t=20$, the fraction of infected in the population is growing (this is not visible in the figure). But the lockdown policy we simulated is strong enough to cause an immediate reduction in the number of infected. The estimated treatment from the regression is however constant over time, by construction, and hence the predicted effects (the dotted lines) continue to increase until the curve reaches its peak in the untreated units. This is the case both when spatial-heterogeneity is controlled for (Figure \ref{fig:prediction_nobias}) and when it is not (Figure \ref{fig:prediction_withbias}). In this second case, furthermore, the prediction overstates true infection levels. This is because, in our simulations, treatment in cities with low density is used to predict treatment in cities with high density, (for which we don't use treatment data in the estimation). In such cities, the ``true'' effect of the policy would be larger (in absolute terms), which is why the dotted line in Figure \ref{fig:prediction_nobias} is lower than the dotted line in Figure \ref{fig:prediction_withbias}.

Finally, we simulate a more realistic scenario where the data includes cities of various densities and various treatment dates.
We report results in Table \ref{tab:est_policy}. The table reports the ``true'' average treatment coefficient, the one we would obtain with data that contain an untreated city for each differently treated one, and the estimated coefficient with partial data.\footnote{We report two specifications, one where we did not include density as a regressor, and another in which we included density and density interacted with treatment.} 
As to be expected, prediction errors are large. \newline 

We conclude with some methodological suggestions for the empirical analysis of epidemiological data across cities and time and for the study of the effects of NPI's in this context, as they are implied by our study in this paper. Spatial and time heterogeneity render the reduced-form analysis of these data  problematic. Short of directly estimating structural models, simulated data can be exploited in an integrated structural/reduced-form empirical analysis.

\begin{table}[t!]
\centering
\caption{
    Coefficients of treatment parameter estimates
} \label{tab:est_policy}

\begin{tabular}{lcccc}
\toprule
&
\multicolumn{2}{c}{Treated}
&
\multicolumn{2}{c}{Treated\#Density}
\\
Outcome
&
True
&
Estimated
&
True
&
Estimated
 \\ \midrule   \\ \multicolumn{5}{c}{Without behavioral responses} \\ \midrule 
\multirow{2}{*}{Infected}&-383.605&-323.088&&\\
 &1006.816&214.176&-1390.421&-620.278\\
\multirow{2}{*}{Contacts}&-0.250&-0.250&&\\
 &-0.250&-0.250&-0.000&-0.000\\
\multirow{2}{*}{Growth rate}&-0.004&-0.009&&\\
 &0.249&0.258&-0.253&-0.306\\
 \\ \multicolumn{5}{c}{With behavioral responses} \\ \midrule 
\multirow{2}{*}{Infected}&-324.242&-419.158&&\\
 &360.378&-319.468&-684.619&-114.588\\
\multirow{2}{*}{Contacts}&-0.197&-0.172&&\\
 &-0.262&-0.221&0.065&0.056\\
\multirow{2}{*}{Growth rate}&-0.013&-0.014&&\\
 &0.171&0.168&-0.184&-0.208\\
\bottomrule \end{tabular}

\bigskip

\caption*{\footnotesize \normalfont \emph{Note:} Estimates of the ``treatment" parameter $\delta$ in (\ref{ns}) and of the interaction of treatment with density using data simulated from the Spatial-SIR models of Figure \ref{fig:est_densitybetas}, with densities 0.5, 1, and 1.5. The ``True'' column reports estimates obtained from data with cities that are never treated, treated at date $t>=15$, and treated at date $t>=40$, for all density levels. The
``Estimated'' column reports estimates obtained from data including only cities with density 1 that are never treated, cities with density 0.5 that are treated at $t>=40$, and cities with density 1.5 that are treated at $t>=15$. For each outcome we report coefficients obtained from a specification without controls for density, and one with controls for density and density interacted with treatment. We omit standard errors, which can be made arbitrarily small by increasing the number of simulated cities.}
\end{table}

Simulated data obtained from structural models, accounting for the relevant spatial and  behavioral characteristics, are in principle very useful to indicate the sign and the size of prediction errors, as our empirical exercises make clear, e.g.,  when comparing the prediction errors in Figure \ref{fig:prediction_nobias} and \ref{fig:prediction_withbias}.  

More specifically, simulated data obtained from structural models are  useful to capture spatial and time heterogeneity flexibly. 
For instance, when estimating Equation \ref{ns} to predict active cases, breaking the sample to distinguish  earlier dates when cases are mostly growing in untreated cities from later ones when they are decreasing appears of first order importance to reduce time-heterogeneity. Similarly, geographic units of analysis should be chosen so that density and other geographic characteristics are relatively homogeneous.\footnote{For this reason, empirical analyses with data across  countries involve additional concerns with respect to data across cities.} But when this is not the case, the sample can be broken by density (and other relevant geographical characteristics) to better control for spatial heterogeneity.   
In these cases, while identifying the natural breaking points in the data may be difficult, simulations can be used to facilitate this procedure, for example, to identify which geographic variables are  most relevant, in order to induce a  parsimonious use of fixed effects and controls in Equation \ref{ns} and capture more of the structure of the dynamics of the epidemic while avoiding overfitting the data.

\section{Conclusions} 

We study the effects of several stylized spatial factors identifying the fundamental role of local interaction and matching frictions as a determinant of the dynamics of epidemic. We highlight important implications for empirical studies on the diffusion of an epidemic, providing a framework for disentangling the effects of local interactions/matching frictions, behavioral responses of risk-averse agents, and policy interventions. 

\newpage
\bibliographystyle{aer}
\bibliography{covid}

@techreport{gans2020,
 title = "The Economic Consequences of $\hat{R}=1$: Towards a Workable Behavioural Epidemiological Model of Pandemics",
 author = "Gans, Joshua S",
 institution = "National Bureau of Economic Research",
 type = "Working Paper",
 series = "Working Paper Series",
 number = "27632",
 year = "2020",
 month = "July",
 doi = {10.3386/w27632},
 URL = "http://www.nber.org/papers/w27632",
}

@techreport{gupta2020effects,
	Author = {Gupta, Sumedha and Montenovo, Laura and Nguyen, Thuy D and Rojas, Felipe Lozano and Schmutte, Ian M and Simon, Kosali I and Weinberg, Bruce A and Wing, Coady},
	Institution = {National Bureau of Economic Research},
	Title = {Effects of social distancing policy on labor market outcomes},
	Year = {2020}}

@techreport{goolsbee2020fear,
	Author = {Goolsbee, Austan and Syverson, Chad},
	Institution = {National Bureau of Economic Research},
	Title = {Fear, lockdown, and diversion: comparing drivers of pandemic economic decline 2020},
	Year = {2020}}

@misc{aguirre,
	Author = {Aguirregabiria, Victor and Gu, Jiaying and Luo, Yao and Mira, Pedro},
	Howpublished = {CEPR Discussion Paper No. DP14750},
	Title = {A Dynamic Structural Model of Virus Diffusion and Network Production: A First Report},
	Year = {2020}}

@article{Chernozhukov2020,
	Author = {Chernozhukov, Victor and Kasahara, Hiroyuki and Schrimpf, Paul},
	Doi = {10.1101/2020.05.27.20115139},
	Elocation-Id = {2020.05.27.20115139},
	Eprint = {https://www.medrxiv.org/content/early/2020/05/30/2020.05.27.20115139.full.pdf},
	Journal = {medRxiv},
	Publisher = {Cold Spring Harbor Laboratory Press},
	Title = {Causal Impact of Masks, Policies, Behavior on Early Covid-19 Pandemic in the U.S.},
	Url = {https://www.medrxiv.org/content/early/2020/05/30/2020.05.27.20115139},
	Year = {2020}}

@misc{ellison,
	Author = {Ellison, Glenn},
	Doi = {10.3386/w27373},
	Howpublished = {National Bureau of Economic Research Working Paper 27373},
	Month = {June},
	Number = {27373},
	Series = {Working Paper Series},
	Title = {Implications of Heterogeneous SIR Models for Analyses of COVID-19},
	Type = {Working Paper},
	Url = {http://www.nber.org/papers/w27373},
	Year = {2020}}

@article{hsiang2020effect,
	Author = {Hsiang, Solomon and Allen, Daniel and Annan-Phan, Sebastien and Bell, Kendon and Bolliger, Ian and Chong, Trinetta and Druckenmiller, Hannah and Hultgren, Andrew and Huang, Luna Yue and Krasovich, Emma and others},
	Journal = {Nature},
	Title = {The effect of large-scale anti-contagion policies on the coronavirus (covid-19) pandemic},
	Year = {2020}}

@misc{fernandez2020estimating,
	Author = {Fernandez-Villaverde, Jesus and Jones, Charles},
	Howpublished = {mimeo},
	Title = {Estimating and Simulating a SIRD Model of COVID-19},
	Year = {2020}}

@article{chinazzi2020effect,
	Author = {Chinazzi, Matteo and Davis, Jessica T and Ajelli, Marco and Gioannini, Corrado and Litvinova, Maria and Merler, Stefano and y Piontti, Ana Pastore and Mu, Kunpeng and Rossi, Luca and Sun, Kaiyuan and others},
	Journal = {Science},
	Number = {6489},
	Pages = {395--400},
	Publisher = {American Association for the Advancement of Science},
	Title = {The effect of travel restrictions on the spread of the 2019 novel coronavirus (COVID-19) outbreak},
	Volume = {368},
	Year = {2020}}

@article{balcan2010modeling,
	Author = {Balcan, Duygu and Gon{\c{c}}alves, Bruno and Hu, Hao and Ramasco, Jos{\'e} J and Colizza, Vittoria and Vespignani, Alessandro},
	Journal = {Journal of computational science},
	Number = {3},
	Pages = {132--145},
	Publisher = {Elsevier},
	Title = {Modeling the spatial spread of infectious diseases: The GLobal Epidemic and Mobility computational model},
	Volume = {1},
	Year = {2010}}

@techreport{alfaro2020social,
	Author = {Alfaro, Laura and Faia, Ester and Lamersdorf, Nora and Saidi, Farzad},
	Institution = {National Bureau of Economic Research},
	Title = {Social Interactions in Pandemics: Fear, Altruism, and Reciprocity},
	Year = {2020}}

@misc{toxvaerd2020equilibrium,
	Author = {Toxvaerd, FMO},
	Howpublished = {Faculty of Economics, University of Cambridge},
	Title = {Equilibrium social distancing},
	Year = {2020}}

@article{balcan2009multiscale,
	Author = {Balcan, Duygu and Colizza, Vittoria and Gon{\c{c}}alves, Bruno and Hu, Hao and Ramasco, Jos{\'e} J and Vespignani, Alessandro},
	Journal = {Proceedings of the National Academy of Sciences},
	Number = {51},
	Pages = {21484--21489},
	Publisher = {National Acad Sciences},
	Title = {Multiscale mobility networks and the spatial spreading of infectious diseases},
	Volume = {106},
	Year = {2009}}

@article{verelst2016behavioural,
	Author = {Verelst, Frederik and Willem, Lander and Beutels, Philippe},
	Journal = {Journal of The Royal Society Interface},
	Number = {125},
	Pages = {20160820},
	Publisher = {The Royal Society},
	Title = {Behavioural change models for infectious disease transmission: a systematic review (2010--2015)},
	Volume = {13},
	Year = {2016}}

@article{funk2010modelling,
	Author = {Funk, Sebastian and Salath{\'e}, Marcel and Jansen, Vincent AA},
	Journal = {Journal of the Royal Society Interface},
	Number = {50},
	Pages = {1247--1256},
	Publisher = {The Royal Society},
	Title = {Modelling the influence of human behaviour on the spread of infectious diseases: a review},
	Volume = {7},
	Year = {2010}}

@article{bethune2020covid,
	Author = {Bethune, Zachary A and Korinek, Anton},
	Journal = {National Bureau of Economic Research Working Paper},
	Title = {Covid-19 infection externalities: Trading off lives vs. livelihoods},
	Year = {2020}}

@techreport{acemoglu2020testing,
	Author = {Acemoglu, Daron and Makhdoumi, Ali and Malekian, Azarakhsh and Ozdaglar, Asuman},
	Institution = {National Bureau of Economic Research},
	Title = {Testing, Voluntary Social Distancing and the Spread of an Infection},
	Year = {2020}}

@book{liggett2012interacting,
	Author = {Liggett, Thomas Milton},
	Publisher = {Springer Science \& Business Media},
	Title = {Interacting particle systems},
	Volume = {276},
	Year = {2012}}

@article{kindermann1980american,
	Author = {Kindermann, R and Snell, JL},
	Journal = {Markov random fields and their applications},
	Title = {American Mathematical Society},
	Year = {1980}}

@inproceedings{keppo2020behavioral,
	Author = {Keppo, Juusi and Kudlyak, Marianna and Quercioli, Elena and Smith, Lones and Wilson, Andrea},
	Booktitle = {Virtual Macro Seminar},
	Title = {The behavioral SIR model, with applications to the Swine Flu and COVID-19 pandemics},
	Year = {2020}}

@misc{acemoglu2020multi,
	Author = {Acemoglu, Daron and Chernozhukov, Victor and Werning, Iv{\'a}n and Whinston, Michael D},
	Howpublished = {National Bureau of Economic Research},
	Title = {A Multi-Risk SIR Model with Optimally Targeted Lockdown},
	Year = {2020}}

@article{goenka2012infectious,
	Author = {Goenka, Aditya and Liu, Lin},
	Journal = {Economic Theory},
	Number = {1},
	Pages = {125--149},
	Publisher = {Springer},
	Title = {Infectious diseases and endogenous fluctuations},
	Volume = {50},
	Year = {2012}}

@article{fenichel2013economic,
	Author = {Fenichel, Eli P},
	Journal = {Journal of health economics},
	Number = {2},
	Pages = {440--451},
	Publisher = {Elsevier},
	Title = {Economic considerations for social distancing and behavioral based policies during an epidemic},
	Volume = {32},
	Year = {2013}}

@article{geoffard1996rational,
	Author = {Geoffard, Pierre-Yves and Philipson, Tomas},
	Journal = {International economic review},
	Pages = {603--624},
	Publisher = {JSTOR},
	Title = {Rational epidemics and their public control},
	Year = {1996}}

@article{eubank2004modelling,
	Author = {Eubank, Stephen and Guclu, Hasan and Kumar, VS Anil and Marathe, Madhav V and Srinivasan, Aravind and Toroczkai, Zoltan and Wang, Nan},
	Journal = {Nature},
	Number = {6988},
	Pages = {180--184},
	Publisher = {Nature Publishing Group},
	Title = {Modelling disease outbreaks in realistic urban social networks},
	Volume = {429},
	Year = {2004}}

@article{kaplan2020gianluca,
	Author = {Kaplan, Greg and Moll, Ben and Violante, Gianluca},
	Journal = {University of Chicago},
	Title = {Pandemics According to HANK},
	Year = {2020}}

@misc{atkeson2020will,
	Author = {Atkeson, Andrew},
	Howpublished = {National Bureau of Economic Research},
	Title = {What will be the economic impact of COVID-19 in the US? Rough estimates of disease scenarios},
	Year = {2020}}

@misc{eichenbaum2020macroeconomics,
	Author = {Eichenbaum, Martin S and Rebelo, Sergio and Trabandt, Mathias},
	Howpublished = {National Bureau of Economic Research},
	Title = {The macroeconomics of epidemics},
	Year = {2020}}

@misc{bognanni2020economic,
	Author = {Bognanni, Mark and Hanley, Doug and Kolliner, Daniel and Mitman, Kurt},
	Howpublished = {mimeo, IIES},
	Title = {Economic Activity and COVID-19 Transmission: Evidence from an Estimated Economic-Epidemiological Model},
	Year = {2020}}

@techreport{fajgelbaum2020optimal,
	Author = {Fajgelbaum, Pablo and Khandelwal, Amit and Kim, Wookun and Mantovani, Cristiano and Schaal, Edouard},
	Institution = {National Bureau of Economic Research},
	Title = {Optimal lockdown in a commuting network},
	Year = {2020}}

@techreport{argente2020cost,
	Author = {Argente, David O and Hsieh, Chang-Tai and Lee, Munseob},
	Institution = {National Bureau of Economic Research},
	Title = {The Cost of Privacy: Welfare Effect of the Disclosure of COVID-19 Cases},
	Year = {2020}}

@misc{cua2020structural,
	Author = {Cu$\tilde{n}$at, Alejandro and Zymek, Robert},
	Howpublished = {CESifo Working Paper},
	Title = {The (Structural) Gravity of Epidemics},
	Year = {2020}}

@misc{BisinMoroRat2020,
	Author = {Bisin, Alberto and Moro, Andrea},
	Howpublished = {mimeo, NYU},
	Title = {Notes on Rational Forward Looking SIR},
	Year = {2020}}

@article{britton2020mathematical,
	Author = {Britton, Tom and Ball, Frank and Trapman, Pieter},
	Journal = {Science},
	Publisher = {American Association for the Advancement of Science},
	Title = {A mathematical model reveals the influence of population heterogeneity on herd immunity to SARS-CoV-2},
	Year = {2020}}

@article{gomes2020individual,
	Author = {Gomes, M Gabriela M and Aguas, Ricardo and Corder, Rodrigo M and King, Jessica G and Langwig, Kate E and Souto-Maior, Caetano and Carneiro, Jorge and Ferreira, Marcelo U and Penha-Goncalves, Carlos},
	Journal = {medRxiv},
	Publisher = {Cold Spring Harbor Laboratory Press},
	Title = {Individual variation in susceptibility or exposure to SARS-CoV-2 lowers the herd immunity threshold},
	Year = {2020}}

@techreport{antras2020globalization,
	Author = {Antr{\`a}s, Pol and Redding, Stephen J and Rossi-Hansberg, Esteban},
	Institution = {Harvard University Working Paper},
	Title = {Globalization and Pandemics},
	Year = {2020}}

@misc{glaeser2020much,
	Author = {Glaeser, Edward L and Gorback, Caitlin S and Redding, Stephen J},
	Howpublished = {mimeo},
	Title = {How Much does Covid-19 Increase with Mobility? Evidence from New York and Four Other U.S. Cities},
	Year = {2020}}

@misc{Azzimontietal2020,
	Author = {Azzimonti, Marina and Fogli, Alessandra and Perri, Fabrizio and Ponder, Mark},
	Howpublished = {Federal Reserve Bank of Minneapolis Staff report 609},
	Title = {Pandemic Control in ECON-EPI Networks},
	Month = {August},
	Year = {2020}
	}

@misc{birge2020controlling,
	Author = {Birge, John R and Candogan, Ozan and Feng, Yiding},
	Howpublished = {University of Chicago, Becker Friedman Institute for Economics Working Paper},
	Number = {2020-57},
	Title = {Controlling Epidemic Spread: Reducing Economic Losses with Targeted Closures},
	Year = {2020}}

@misc{alvarez2020simple,
	Author = {Alvarez, Fernando E and Argente, David and Lippi, Francesco},
	Howpublished = {National Bureau of Economic Research},
	Title = {A simple planning problem for covid-19 lockdown},
	Year = {2020}}

@misc{Desmet2020,
	Author = {Desmet, Klaus and Wacziarg, Romain},
	Howpublished = {National Bureau of Economic Research},
	Title = {Understanding spatial variation in COVID-19 across the United States},
	Year = {2020}}

@article{kermack1927contribution,
	Author = {Kermack, William Ogilvy and McKendrick, Anderson G},
	Journal = {Proceedings of the royal society of london. Series A, Containing papers of a mathematical and physical character},
	Number = {772},
	Pages = {700--721},
	Publisher = {The Royal Society London},
	Title = {A contribution to the mathematical theory of epidemics},
	Volume = {115},
	Year = {1927}}

@article{greenwood2019equilibrium,
	Author = {Greenwood, Jeremy and Kircher, Philipp and Santos, Cezar and Tertilt, Mich{\`e}le},
	Journal = {Econometrica},
	Number = {4},
	Pages = {1081--1113},
	Publisher = {Wiley Online Library},
	Title = {An equilibrium model of the African HIV/AIDS epidemic},
	Volume = {87},
	Year = {2019}}

@misc{ferguson2020report,
	Author = {Ferguson, Neil and Laydon, Daniel and Nedjati Gilani, Gemma and Imai, Natsuko and Ainslie, Kylie and Baguelin, Marc and Bhatia, Sangeeta and Boonyasiri, Adhiratha and Cucunuba Perez, ZULMA and Cuomo-Dannenburg, Gina and others},
	Howpublished = {Imperial College London},
	Title = {Imperial College COVID-19 Response Team: Impact of non-pharmaceutical interventions (NPIs) to reduceCOVID-19 mortality and healthcare demand},
	Year = {2020}}

@article{Mizumoto_2020,
	Author = {Mizumoto, Kenji and Kagaya, Katsushi and Zarebski, Alexander and Chowell, Gerardo},
	Issn = {1560-7917},
	Journal = {Eurosurveillance},
	Month = {Mar},
	Number = {10},
	Publisher = {European Centre for Disease Control and Prevention (ECDC)},
	Title = {Estimating the asymptomatic proportion of coronavirus disease 2019 (COVID-19) cases on board the Diamond Princess cruise ship, Yokohama, Japan, 2020},
	Volume = {25},
	Year = {2020}}

@article{Mossong_2008,
	Author = {Mossong, Jo{\"e}l and Hens, Niel and Jit, Mark and Beutels, Philippe and Auranen, Kari and Mikolajczyk, Rafael and Massari, Marco and Salmaso, Stefania and Tomba, Gianpaolo Scalia and Wallinga, Jacco and et al.},
	Editor = {Riley, StevenEditor},
	Issn = {1549-1676},
	Journal = {PLoS Medicine},
	Month = {Mar},
	Number = {3},
	Pages = {e74},
	Publisher = {Public Library of Science (PLoS)},
	Title = {Social Contacts and Mixing Patterns Relevant to the Spread of Infectious Diseases},
	Volume = {5},
	Year = {2008}}

@article{giannone,
	Author = {Giannone, Elisa and Paix{\~a}o, Nuno and Pang, Xinle
},
	Journal = {Covid Economics: Vetted and Real-Time Papers},
	Month = {October},
	Pages = {68-95},
	Title = {The Geography of Pandemic Containment},
	Volume = {52},
	Year = {2020}}

@article{huang2020clinical,
	Author = {Huang, Chaolin and Wang, Yeming and Li, Xingwang and Ren, Lili and Zhao, Jianping and Hu, Yi and Zhang, Li and Fan, Guohui and Xu, Jiuyang and Gu, Xiaoying and others},
	Journal = {The lancet},
	Number = {10223},
	Pages = {497--506},
	Publisher = {Elsevier},
	Title = {Clinical features of patients infected with 2019 novel coronavirus in Wuhan, China},
	Volume = {395},
	Year = {2020}}

@article{remuzzi2020covid,
	Author = {Remuzzi, Andrea and Remuzzi, Giuseppe},
	Journal = {The Lancet},
	Publisher = {Elsevier},
	Title = {COVID-19 and Italy: what next?},
	Year = {2020}}

@article{zhang2020estimation,
	Author = {Zhang, Sheng and Diao, MengYuan and Yu, Wenbo and Pei, Lei and Lin, Zhaofen and Chen, Dechang},
	Journal = {International Journal of Infectious Diseases},
	Pages = {201--204},
	Publisher = {Elsevier},
	Title = {Estimation of the reproductive number of novel coronavirus (COVID-19) and the probable outbreak size on the Diamond Princess cruise ship: A data-driven analysis},
	Volume = {93},
	Year = {2020}}

@article{paules2020coronavirus,
	Author = {Paules, Catharine I and Marston, Hilary D and Fauci, Anthony S},
	Journal = {Jama},
	Number = {8},
	Pages = {707--708},
	Publisher = {American Medical Association},
	Title = {Coronavirus infections?more than just the common cold},
	Volume = {323},
	Year = {2020}}

@article{kermack1932contributions,
	Author = {Kermack, William Ogilvy and McKendrick, Anderson G},
	Journal = {Proceedings of the Royal Society of London. Series A, containing papers of a mathematical and physical character},
	Number = {834},
	Pages = {55--83},
	Publisher = {The Royal Society London},
	Title = {Contributions to the mathematical theory of epidemics. II. -The problem of endemicity},
	Volume = {138},
	Year = {1932}}

@article{neumeyer2020clase,
	Author = {Neumeyer, Pablo Andres},
	Journal = {Author's website, Class notes, Universidad Di Tella},
	Title = {Clase especial de epidemilogia},
	Year = {2020}}

@article{Moll2020,
	Author = {Moll, Ben},
	Journal = {Author's website, Princeton},
	Title = {Lockdowns in SIR models},
	Year = {2020}}

@misc{jarosch2020internal,
	Author = {Jarosch, Gregor and Farboodi, Maryam and Shimer, Robert},
	Title = {Internal and External Effects of Social Distancing in a Pandemic},
	Year = {2020}}

@article{fang2020human,
	Author = {Fang, Hanming and Wang, Long and Yang, Yang},
	Journal = {National Bureau of Economic Research},
	Title = {Human mobility restrictions and the spread of the novel coronavirus (2019-ncov) in china},
	Year = {2020}}

@article{tome2010critical,
	Author = {Tom{\'e}, T{\^a}nia and Ziff, Robert M},
	Journal = {Physical Review E},
	Number = {5},
	Pages = {051921},
	Publisher = {APS},
	Title = {Critical behavior of the susceptible-infected-recovered model on a square lattice},
	Volume = {82},
	Year = {2010}}

@article{herben42hethcote,
	Author = {Hethcote, Herben W.},
	Journal = {SIAM Review, 2O00},
	Number = {4},
	Pages = {599-653},
	Title = {The mathematics of infectious diseases},
	Volume = {42},
	Year = {2000}}

@article{glaeser2001measuring,
	Author = {Glaeser, Edward and Scheinkman, Jos{\'e}},
	Journal = {Social dynamics},
	Pages = {83--132},
	Publisher = {Boston, MA: MIT Press},
	Title = {Measuring social interactions},
	Year = {2001}}

@incollection{blume2011identification,
	Author = {Blume, Lawrence E and Brock, William A and Durlauf, Steven N and Ioannides, Yannis M},
	Booktitle = {Benhabib, J, A. Bisin, and M. Jackson (eds.), Handbook of social economics},
	Pages = {853--964},
	Publisher = {Elsevier},
	Title = {Identification of social interactions},
	Volume = {1},
	Year = {2011}}

@misc{ozgur2011dynamic,
	Author = {{\"O}zg{\"u}r, Onur and Bisin, Alberto and Bramoull{\'e}, Yann},
	Howpublished = {Melbourne Business School working paper},
	Title = {Dynamic linear economies with social interactions},
	Year = {2019}}

@article{conley2007estimating,
	Author = {Conley, Timothy G and Topa, Giorgio},
	Journal = {Journal of Econometrics},
	Number = {1},
	Pages = {282--303},
	Publisher = {Elsevier},
	Title = {Estimating dynamic local interactions models},
	Volume = {140},
	Year = {2007}}

@misc{bisinmoro2020,
	Author = {Bisin, Alberto and Moro, Andrea},
	Journal = {Working paper, NYU and Vanderbilt},
	Title = {Spatial-SIR with Network Structure and Behavior: Lockdown Policies and the Lucas Critique},
	Year = {2020},
	howpublished = {Working paper, NYU and Vanderbilt University}
	}

@article{mangrum2020college,
	Author = {Mangrum, Daniel and Niekamp, Paul},
	Journal = {Available at SSRN 3606811},
	Title = {College Student Contribution to Local COVID-19 Spread: Evidence from University Spring Break Timing},
	Year = {2020}}

@misc{courtemanche2020did,
	Author = {Courtemanche, Charles J and Garuccio, Joseph and Le, Anh and Pinkston, Joshua C and Yelowitz, Aaron},
	Howpublished = {Working paper},
	Title = {Did Social-Distancing Measures in Kentucky Help to Flatten the COVID-19 Curve?},
	Year = {2020}}

@article{grassberger1983critical,
	Author = {Grassberger, Peter},
	Journal = {Mathematical Biosciences},
	Number = {2},
	Pages = {157--172},
	Publisher = {Elsevier},
	Title = {On the critical behavior of the general epidemic process and dynamical percolation},
	Volume = {63},
	Year = {1983}}

@article{allcott2020polarization,
	Author = {Allcott, Hunt and Boxell, Levi and Conway, Jacob and Gentzkow, Matthew and Thaler, Michael and Yang, David Y},
	Journal = {National Bureau of Economic Research Working Paper},
	Title = {Polarization and public health: Partisan differences in social distancing during the Coronavirus pandemic},
	Volume = {w26946},
	Year = {2020}}

@article{bacon2020,
	Author = {Goodman-Bacon, Andrew and Marcus, Jan},
	Journal = {DIW Berlin Discussion Paper},
	Title = {Using Difference-in-Differences to Identify Causal Effects of COVID-19 Policies},
	Year = {2020}}

@article{maloney2020determinants,
	Author = {Maloney, William F and Taskin, Temel},
	Journal = {Covid Economics},
	Month = {May},
	Title = {Determinants of Social Distancing and Economic Activity during COVID-19: A Global View},
	Volume = {Issue 13},
	Year = {2020}}

@article{pepe2020covid,
	Author = {Pepe, Emanuele and Bajardi, Paolo and Gauvin, Laetitia and Privitera, Filippo and Lake, Brennan and Cattuto, Ciro and Tizzoni, Michele},
	Journal = {medRxiv},
	Publisher = {Cold Spring Harbor Laboratory Press},
	Title = {COVID-19 outbreak response: a first assessment of mobility changes in Italy following national lockdown},
	Year = {2020}}

@article{brotherhood2020economic,
	Author = {Brotherhood, Luiz and Kircher, Philipp and Santos, Cezar and Tertilt, Mich{\`e}le},
	Journal = {CEPR Discussion Paper No. DP14695},
	Title = {An economic model of the Covid-19 epidemic: The importance of testing and age-specific policies},
	Year = {2020}}

@article{weitz2020moving,
	Author = {Weitz, Joshua S and Park, Sang Woo and Eksin, Ceyhun and Dushoff, Jonathan},
	Journal = {medRxiv},
	Publisher = {Cold Spring Harbor Laboratory Press},
	Title = {Moving Beyond a Peak Mentality: Plateaus, Shoulders, Oscillations and Other'Anomalous' Behavior-Driven Shapes in COVID-19 Outbreaks},
	Year = {2020}}

@article{Lavezzo-Vo-Study,
	Author = {Lavezzo, Enrico and Franchin, Elisa and Ciavarella, Constanze and Cuomo-Dannenburg, Gina and Barzon, Luisa and Del Vecchio, Claudia and Rossi, Lucia and Manganelli, Riccardo and Loregian, Arianna and Navarin, Nicol{\`o} and Abate, Davide and Sciro, Manuela and Merigliano, Stefano and Decanale, Ettore and Vanuzzo, Maria Cristina and Saluzzo, Francesca and Onelia, Francesco and Pacenti, Monia and Parisi, Saverio and Carretta, Giovanni and Donato, Daniele and Flor, Luciano and Cocchio, Silvia and Masi, Giulia and Sperduti, Alessandro and Cattarino, Lorenzo and Salvador, Renato and Gaythorpe, Katy A.M. and Brazzale, Alessandra R and Toppo, Stefano and Trevisan, Marta and Baldo, Vincenzo and Donnelly, Christl A. and Ferguson, Neil M. and Dorigatti, Ilaria and Crisanti, Andrea},
	Doi = {10.1101/2020.04.17.20053157},
	Elocation-Id = {2020.04.17.20053157},
	Eprint = {https://www.medrxiv.org/content/early/2020/04/18/2020.04.17.20053157.full.pdf},
	Journal = {medRxiv},
	Publisher = {Cold Spring Harbor Laboratory Press},
	Title = {Suppression of COVID-19 outbreak in the municipality of Vo, Italy},
	Url = {https://www.medrxiv.org/content/early/2020/04/18/2020.04.17.20053157},
	Year = {2020}}

@article{duranton-puga,
Author = {Duranton, Gilles and Puga, Diego},
Title = {The Economics of Urban Density},
Journal = {Journal of Economic Perspectives},
Volume = {34},
Number = {3},
Year = {2020},
Month = {August},
Pages = {3-26},
DOI = {10.1257/jep.34.3.3},
URL = {https://www.aeaweb.org/articles?id=10.1257/jep.34.3.3}}

@techreport{couture,
 title = "Measuring Movement and Social Contact with Smartphone Data: A Real-Time Application to COVID-19",
 author = "Couture, Victor and Dingel, Jonathan I and Green, Allison E and Handbury, Jessie and Williams, Kevin R",
 institution = "National Bureau of Economic Research",
 type = "Working Paper",
 series = "Working Paper Series",
 number = "27560",
 year = "2020",
 month = "July",
 doi = {10.3386/w27560},
 URL = "http://www.nber.org/papers/w27560",
}

@misc{chetty2020economic,
  title={The economic impacts of COVID-19: Evidence from a new public database built from private sector data},
  author={Chetty, Raj and Friedman, J and Hendren, Nathaniel and Stepner, Michael},
  year={2020},
  howpublished={Opportunity Insights Working Paper.}
}

@techreport{farboodi2020,
 title = "Internal and External Effects of Social Distancing in a Pandemic",
 author = "Farboodi, Maryam and Jarosch, Gregor and Shimer, Robert",
 institution = "National Bureau of Economic Research",
 type = "Working Paper",
 series = "Working Paper Series",
 number = "27059",
 year = "2020",
 month = "April",
 doi = {10.3386/w27059},
 URL = "http://www.nber.org/papers/w27059",
}

@article{el2012social,
  title={Social network analysis and agent-based modeling in social epidemiology},
  author={El-Sayed, Abdulrahman M and Scarborough, Peter and Seemann, Lars and Galea, Sandro},
  journal={Epidemiologic Perspectives \& Innovations},
  volume={9},
  number={1},
  pages={1},
  year={2012},
  publisher={Springer}
}

@article{bruch2015agent,
  title={Agent-based models in empirical social research},
  author={Bruch, Elizabeth and Atwell, Jon},
  journal={Sociological methods \& research},
  volume={44},
  number={2},
  pages={186--221},
  year={2015},
  publisher={SAGE Publications Sage CA: Los Angeles, CA}
}

@article{dunham2005agent,
  title={An agent-based spatially explicit epidemiological model in MASON},
  author={Dunham, Jill Bigley},
  journal={Journal of Artificial Societies and Social Simulation},
  volume={9},
  number={1},
  year={2005}
}

@article{grefenstette2013fred,
  title={FRED (A Framework for Reconstructing Epidemic Dynamics): an open-source software system for modeling infectious diseases and control strategies using census-based populations},
  author={Grefenstette, John J and Brown, Shawn T and Rosenfeld, Roni and DePasse, Jay and Stone, Nathan TB and Cooley, Phillip C and Wheaton, William D and Fyshe, Alona and Galloway, David D and Sriram, Anuroop and others},
  journal={BMC public health},
  volume={13},
  number={1},
  pages={1--14},
  year={2013},
  publisher={BioMed Central}
}

@inproceedings{hunter2018comparison,
  title={A Comparison of Agent-Based Models and Equation Based Models for Infectious Disease Epidemiology.},
  author={Hunter, Elizabeth and Mac Namee, Brian and Kelleher, John D},
  booktitle={AICS},
  pages={33--44},
  year={2018}
}

@article{hunter2017taxonomy,
  title={A taxonomy for agent-based models in human infectious disease epidemiology},
  author={Hunter, Elizabeth and Mac Namee, Brian and Kelleher, John D},
  journal={Journal of Artificial Societies and Social Simulation},
  volume={20},
  number={3},
  year={2017},
  publisher={JASSS}
}

@article{chinviriyasit2010numerical,
  title={Numerical modelling of an SIR epidemic model with diffusion},
  author={Chinviriyasit, Settapat and Chinviriyasit, Wirawan},
  journal={Applied Mathematics and Computation},
  volume={216},
  number={2},
  pages={395--409},
  year={2010},
  publisher={Elsevier}
}

@article{wu2017epidemic,
  title={Epidemic waves of a spatial SIR model in combination with random dispersal and non-local dispersal},
  author={Wu, Chufen and Yang, Yong and Zhao, Qianyi and Tian, Yanling and Xu, Zhiting},
  journal={Applied Mathematics and Computation},
  volume={313},
  pages={122--143},
  year={2017},
  publisher={Elsevier}
}

@book{billari2012agent,
  title={Agent-based computational demography: Using simulation to improve our understanding of demographic behaviour},
  author={Billari, Francesco C and Prskawetz, Alexia},
  year={2012},
  publisher={Springer Science \& Business Media}
}

@article{yilmazkuday2020covid,
  title={COVID-19 and unequal social distancing across demographic groups},
  author={Yilmazkuday, Hakan},
  journal={Regional Science Policy \& Practice},
  volume={12},
  number={6},
  pages={1235--1248},
  year={2020},
  publisher={Wiley Online Library}
}

@article{CALLAWAY2020,
title = {Difference-in-Differences with multiple time periods},
journal = {Journal of Econometrics},
year = {2020},
issn = {0304-4076},
doi = {https://doi.org/10.1016/j.jeconom.2020.12.001},
url = {https://www.sciencedirect.com/science/article/pii/S0304407620303948},
author = {Brantly Callaway and Pedro H.C. Sant'Anna}
}

@misc{callaway2021policy,
      title={Policy Evaluation during a Pandemic}, 
      author={Brantly Callaway and Tong Li},
      year={2021},
      eprint={2105.06927},
      howpublished={arXiv working paper},
      primaryClass={econ.EM}
}

@techreport{allcott2020explains,
  title={What Explains Temporal and Geographic Variation in the Early US Coronavirus Pandemic?},
  author={Allcott, Hunt and Boxell, Levi and Conway, Jacob C and Ferguson, Billy A and Gentzkow, Matthew and Goldman, Benny},
  year={2020},
  institution={National Bureau of Economic Research}
}

@techreport{bartik2020measuring,
  title={Measuring the labor market at the onset of the COVID-19 crisis},
  author={Bartik, Alexander W and Bertrand, Marianne and Lin, Feng and Rothstein, Jesse and Unrath, Matt},
  year={2020},
  institution={National Bureau of Economic Research}
}

@article{dave2020jue,
  title={JUE Insight: Were urban cowboys enough to control COVID-19? Local shelter-in-place orders and coronavirus case growth},
  author={Dave, Dhaval and Friedson, Andrew and Matsuzawa, Kyutaro and Sabia, Joseph J and Safford, Samuel},
  journal={Journal of urban economics},
  pages={103294},
  year={2020},
  publisher={Elsevier}
}

@article{courtemanche2020strong,
  title={Strong Social Distancing Measures In The United States Reduced The COVID-19 Growth Rate: Study evaluates the impact of social distancing measures on the growth rate of confirmed COVID-19 cases across the United States.},
  author={Courtemanche, Charles and Garuccio, Joseph and Le, Anh and Pinkston, Joshua and Yelowitz, Aaron},
  journal={Health Affairs},
  volume={39},
  number={7},
  pages={1237--1246},
  year={2020}
}

@article{brady2020fragmented,
  title={The Fragmented United States of America: The impact of scattered lock-down policies on country-wide infections},
  author={Brady, Ryan Robert and Insler, Mike and Rothert, Jacek},
  journal={COVID Economics (43)},
  pages={42--94},
  year={2020}
}

@article{gapen2020assessing,
  title={Assessing the effectiveness of alternative measures to slow the spread of COVID-19 in the United States},
  author={Gapen, Michael and Millar, Jonathan and Blerina, U and Sriram, Pooja},
  journal={Covid Economics},
  volume={40},
  pages={46--75},
  year={2020}
}

@misc{juranek2020effect,
  title={The effect of social distancing measures on the demand for intensive care: Evidence on COVID-19 in Scandinavia},
  author={Juranek, Steffen and Zoutman, Floris},
  year={2020},
  howpublished={CESifo Working Paper}
}

@article{kong2020disentangling,
  title={Disentangling policy effects using proxy data: Which shutdown policies affected unemployment during the COVID-19 pandemic?},
  author={Kong, Edward and Prinz, Daniel},
  journal={Journal of Public Economics},
  volume={189},
  pages={104257},
  year={2020},
  publisher={Elsevier}
}

@article{sears2020we,
  title={Are we\# stayinghome to Flatten the Curve?},
  author={Sears, James and Villas-Boas, J Miguel and Villas-Boas, Vasco and Villas-Boas, Sofia Berto},
  journal={Department of Agricultural and Resource Economics, first version April},
  volume={5},
  year={2020}
}

@techreport{ziedan2020effects,
  title={Effects of state COVID-19 closure policy on non-COVID-19 health care utilization},
  author={Ziedan, Engy and Simon, Kosali I and Wing, Coady},
  year={2020},
  institution={National Bureau of Economic Research}
}

@article{wright2020poverty,
  title={Poverty and economic dislocation reduce compliance with covid-19 shelter-in-place protocols},
  author={Wright, Austin L and Sonin, Konstantin and Driscoll, Jesse and Wilson, Jarnickae},
  journal={Journal of Economic Behavior \& Organization},
  volume={180},
  pages={544--554},
  year={2020},
  publisher={Elsevier}
}

@techreport{crucini2020stay,
  title={Stay-at-Home Orders in a Fiscal Union},
  author={Crucini, Mario J and O'Flaherty, Oscar},
  year={2020},
  institution={National Bureau of Economic Research}
}

@article{borri2020great,
  title={The" Great Lockdown": inactive workers and mortality by Covid-19},
  author={Borri, Nicola and Drago, Francesco and Santantonio, Chiara and Sobbrio, Francesco and others},
  journal={medRxiv},
  year={2020},
  publisher={Cold Spring Harbor Laboratory Press}
}

@article{breen2021distributional,
  title={The distributional impact of Covid-19: Geographic variation in mortality in England},
  author={Breen, Richard and Ermisch, John},
  journal={Demographic Research},
  volume={44},
  pages={397--414},
  year={2021}
}

@misc{borsati2020questioning,
  title={QUESTIONING THE SPATIAL ASSOCIATION BETWEEN THE SPREAD OF COVID-19 AND TRANSIT USAGE IN ITALY},
  author={Borsati, Mattia and Nocera, Silvio and Percoco, Marco},
  year={2020},
  howpublished={Bocconi University Working Paper Series}
}

@misc{wilson2021weather,
  title={Weather, Mobility, and COVID-19: A Panel Local Projections Estimator for Understanding and Forecasting Infectious Disease Spread},
  author={Wilson, Daniel J},
  year={2021},
  howpublished={Federal Reserve Bank of San Francisco}
}

@techreport{glaeser2020learning,
  title={Learning from deregulation: the asymmetric impact of lockdown and reopening on risky behavior during COVID-19},
  author={Glaeser, Edward L and Jin, Ginger Zhe and Leyden, Benjamin T and Luca, Michael},
  year={2020},
  institution={National Bureau of Economic Research}
}

\makeatletter\@input{xx.tex}\makeatother

\end{document}